\documentclass[onefignum,oneabnum]{siamart171218}

\usepackage{amsfonts}							
\usepackage{amsmath,bm}									
\usepackage{graphicx,epstopdf}
\usepackage{subfigure}
\usepackage{algorithm}
\usepackage{algorithmic}

\newcommand{\TheAuthors}{Luchan Zhang and Yang Xiang}

\headers{A new formulation of motion of grain boundaries}{\TheAuthors}

\title{{A new formulation of coupling and sliding motions of grain boundaries based on dislocation structure}\thanks{\funding{This work was partially supported by the Hong Kong Research Grants Council
General Research Fund 16302818.}}}

\author{
  Luchan Zhang and Yang Xiang\thanks{Department of Mathematics, Hong Kong University of Science and Technology, Clear Water Bay, Kowloon, Hong Kong
    (\email{malczhang@ust.hk, maxiang@ust.hk}).}
}

\begin{document}

\maketitle

\begin{abstract}
A continuum model of the two dimensional low angle grain boundary motion and the dislocation structure evolution on the grain boundaries has been developed in Ref.~\cite{ZhangXiang2018157}. The model is based on the motion and reaction of the constituent dislocations of the grain boundaries. The long-range elastic interaction between dislocations is included in the continuum model, and it maintains a stable dislocation structure described by  the Frank's formula for grain boundaries.
In this paper, we develop a new continuum model for the coupling and sliding motions of grain boundaries that avoids the time-consuming calculation of the long-range elastic interaction. In this model, the long-range elastic interaction is replaced by a constraint of the Frank's formula.  The constrained evolution problem in our new continuum model
is further solved by using the projection method. Effects of the coupling and sliding motions in our new continuum model and relationship with the classical motion by curvature model are discussed.
The continuum model is validated by comparisons with discrete dislocation dynamics model and the early continuum model \cite{ZhangXiang2018157} in which the long-range dislocation interaction is explicitly calculated.

\end{abstract}



\section{Introduction}
Grain boundaries are the interfaces between grains with different orientations in  polycrystalline materials \cite{Sutton1995}. The energy and dynamics of grain boundaries play important roles in the mechanical and plastic behaviors of the materials.
These properties of grain boundaries
strongly depend on their underlying microstructures.

The classical grain boundary dynamics models are based upon the motion driven by a capillary force that is proportional to grain boundary mean curvature~\cite{Herring1951,Mullins1956,Sutton1995}. Mathematically, this process is a gradient flow that reduces the total interfacial energy $\int_{\Gamma} \gamma ds$, where $\Gamma$ is the grain boundary, $\gamma$ is the grain boundary energy density, and $s$ is the arclength parameter of the grain boundary, in a two dimensional setting. In this grain boundary dynamics process, the energy density $\gamma$ is fixed, and the total energy is reduced due to the decrease of the grain boundary length. Since the grain boundary energy density $\gamma$ depends on the misorientation angle $\theta$  between the two neighboring grains~\cite{ReadShockley1950,Sutton1995}, this dynamics formulation implies that the misorientation angle $\theta$ is fixed during the grain boundary motion. There are extensive studies in the literature on this motion of grain boundaries by using molecular dynamics or continuum simulations, e.g. \cite{Chenlq1994,Upmanyu1998,Kazaryan2000,liuchun2001,Upmanyu2002,feng2003,Zhang2005,kinderlehrer2006,
du2009,Selim2009,Srolovitz2010,Selim2013,libo2018}.

The grain boundary energy density $\gamma(\theta)$ may also decrease to further reduce the total energy of grain boundaries. In this process, the misorientation angle $\theta$ evolves towards a nearby local energy minimum state. When one grain is embedded into another, this causes rotation of the inner grain. This process is called sliding motion of the grain boundaries, which is a relative rigid-body translation of the grains along the boundary by sliding to reduce the grain boundary energy.
Grain boundary sliding has been observed in experiments and atomistic simulations, and continuum models based on gradient flow with respect to misorientation angle $\theta$ have been developed \cite{Li1962,Shewmon1966,Harris19982623,Kobayashi2000,Upmanyu2006,Selim2016,Liu2019gb,Liu2019gb2}. When the misorientation angle is low, the energy density $\gamma(\theta)$ is an increasing function of misorientation angle $\theta$ and the sliding motion reduces the misorientation angle $\theta$.

 There is a different  mechanism of grain boundary motion in which the grain boundary normal motion induces a tangential translation motion proportionally,  as a result of the geometric constraint that the lattice planes must be continuous across the grain boundary \cite{Li1953223,Cahn20021,Cahn20044887}. This is called the coupling motion of the grain boundaries, and its mechanism was explained by the motion of the constituent dislocations of the low angle grain boundaries whose number is unchanged during the evolution. In the coupling motion, the grain boundary  energy density $\gamma(\theta)$ may increase, unlike  the motion by mean curvature or the sliding motion in which the energy density $\gamma(\theta)$ is constant or decreasing, although the total grain boundary energy $\int_{\Gamma} \gamma ds$ is still decreasing.
Cahn and Taylor \cite{Cahn20044887} formulated the phenomena of the coupling and sliding associated with the grain boundary motion. In their theory, the total tangential velocity $v_{\parallel}$ is the superposition of the tangential velocities generated by these two effect: $v_{\parallel}=\beta v_{\rm n}+v_{\rm s}$, where $\beta v_{\rm n}$ is the tangential velocity induced by the coupling effect,  $v_{\rm n}$ is the grain boundary normal velocity,  $\beta$ is  the coupling parameter, and $v_{\rm s}$ is the tangential velocity produced by the sliding effect.
They also discussed the rotation of circular grain boundaries \cite{Cahn20044887} and proposed a generalized theory for unsymmetrical grain boundaries based on mass transfer by diffusion \cite{Taylor2007493}.
Atomistic simulations and experiments have been performed to validate the theory of Cahn and Taylor \cite{Cahn20021,Cahn20044887,Cahn20064953,Molodov2007,Trautt20122407,Wu2012407,Voorhees2016264,Voorhees2017,Voigt2018}. A special case of cancellation of coupling and sliding effects during the motion of grain boundaries has been showed by a dislocation model and experimental observations \cite{Rath2007}, and the misorientation angle $\theta$ is unchanged during such motion of grain boundaries.

Previously we have derived a continuum model  for the dynamics of low angle grain boundaries incorporating the coupling and sliding motions \cite{ZhangXiang2018157}, from the discrete dislocation dynamics model. The continuum model is based on the theory that a low angle grain boundary consists of regular arrays of dislocations (line defects) \cite{ReadShockley1950,HL,Sutton1995}.
The continuum model includes evolution equations for both the motion of grain boundaries and the evolution of dislocation structure on the grain boundaries, and its underlying mechanisms are the motion and reaction of the constituent dislocations of the grain boundaries.
This model is able to describe the increase of energy density (misorientation angle) due to the coupling motion, the decrease of energy density (misorientation angle) due to the sliding motion, and in general, a combined effect when both of these two tangential motion mechanisms are present. This model can be considered as a generalization of the  Cahn-Taylor theory \cite{Cahn20044887} by  incorporating detailed formulas of the driving forces for the normal and tangential grain boundary motions that depend on the constituent dislocations, their Burgers vectors, the grain boundary shape and the  shape change, and the applied stress.
A key ingredient of this continuum model is the long-range elastic interaction between the constituent dislocations of the grain boundaries, which maintains stable dislocation structures on the grain boundaries described by the Frank's formula (which is an equilibrium condition equivalent to cancellation of the far-field elastic fields). This long-range dislocation interaction takes the form of an integral over the entire grain boundaries, whose  calculation is time-consuming  in  numerical simulations.  There are also some crystal plasticity models  available in the literature that include shear-coupled grain boundary motion \cite{Admal2018,Ask2018}. In these models, densities of  geometric necessary dislocations (net dislocations) rather than the actual dislocation structures were used in the phase field framework \cite{Kobayashi2000} with phenomenological energy of the constituent dislocations.

In this paper, we develop a new continuum model for the coupling and sliding motions of low angle grain boundaries based on their dislocation structure that avoids calculation of the long-range elastic interaction of dislocations. In this model, the long-range elastic interaction in our early continuum model proposed in Ref.~\cite{ZhangXiang2018157} is replaced by a constraint of the Frank's formula. This approximation is based on the observation that the long-range interaction between the grain boundary dislocations is so strong that an equilibrium state described by the Frank's formula is quickly reached during the evolution of the grain boundaries. The equations in the constraint of the Frank's formula are local, thus the calculation of the long-range elastic interaction is avoided. We further solve the constrained evolution problem in our new continuum model  by using the projection method \cite{Chong2012}. Recall that in a project method for constrained evolution problem, the first step is to evolve the system without the constraint, and the second step is to project the obtained new state in the first step into the space that satisfies the constraint. We have found an analytical local formula for the velocity of the grain boundaries after these two steps of the projection method, thus explicit projection procedure is not necessary in our new continuum model.
The result of the projection gives an evolution equation of the misorietation angle $\theta$, in addition to the (adjusted) evolutions of grain boundary and dislocation densities on the grain boundary.
This new, local continuum model describes the coupling and sliding motions of low angle grain boundaries in a simpler form, which is more computationally efficient and more convenient for the analyses of the model and the associated properties of the coupling and sliding motions.

The continuum model is validated by comparisons with discrete dislocation dynamics model and our early continuum model \cite{ZhangXiang2018157} in which the long-range dislocation interaction is explicitly calculated. In particular, under the pure coupling motion with dislocation conservation, the velocity formulation in the new continuum model gives a shape-preserving shrinking motion for a grain boundary in two dimensions, which agrees with the predictions of our early continuum model \cite{ZhangXiang2018157} in which long-range dislocation interaction is explicitly calculated and the  continuum model in Ref.~\cite{Taylor2007493} based on mass transfer via surface diffusion. As our early continuum model in Ref.~\cite{ZhangXiang2018157}, the evolution of dislocation densities at each point on the grain boundary in the new continuum model (due to motion and reaction of the dislocations) enables a description of the accompanied shape changes with the evolution of the misorientation angle during the sliding motion. Except in Ref.~\cite{Taylor2007493}, such accompanied shape changes were not considered in most of the available continuum models for the sliding motion, in which evolution (decrease) of the  misorientation angle and evolution of the grain boundary are independent of each other. For a special case of motion and reaction of the constituent dislocations of the grain boundaries, our continuum model recovers the classical motion by curvature model of the grain boundaries.

The rest of the paper is organized as follows. In Sec.~\ref{sec:settings}, we describe the two dimensional settings of grain boundaries with dislocation structures. In Sec~\ref{sec:review}, the continuum model for grain boundary coupling and sliding motions with long-range dislocation interaction presented in Ref.~\cite{ZhangXiang2018157} (formulation F0) is reviewed.
In Sec.~\ref{sec:new_model}, we present a new continuum model in the form of constrained evolution (F1) in which the long-range force is replaced by the constraint of Frank's formula, and further derive a continuum formulation without constraint (F2) by solving the constrained evolution in formulation (F1) using the projection method.
In Sec.~\ref{sec:numerical}, we validate our continuum formulation (F2) by comparisons with results of the continuum model with the long-range  interaction (F0) and discrete dislocation dynamics simulations.

\section{Grain boundaries with dislocation structure in two dimensions}\label{sec:settings}

We consider a two dimensional grain boundary $\Gamma$ in the $xy$ plane. The inner grain has a misorientation angle $\theta$ relative to the outer grain.
The rotation axis of the grain boundary is parallel to the $z$ axis, therefore the grain boundary is a pure tilt boundary. We focus on the low angle grain boundary that consists of arrays of dislocations~\cite{ReadShockley1950}. Assume that on the grain boundary, there are $J$ dislocation arrays corresponding to $J$ different Burgers vectors $\mathbf{b}^{(j)}=(b_1^{(j)},b_2^{(j)})$ with length $b^{(j)}=\|\mathbf b^{(j)}\|$ , $j=1,2,\cdots,J$, respectively.   All the dislocations are parallel to the $z$ axis, i.e., they are points in the $xy$ plane. The densities (per unit length)  of these constituent dislocations are $\rho^{(j)}$, $j=1,2,\cdots,J$. We allow the dislocation density $\rho^{(j)}$ to be negative to also include dislocations with the same Burgers vector but opposite line direction. Examples of such two-dimensional cases are  Burgers vectors $\mathbf{b}^{(1)} = (b,0)$, $\mathbf{b}^{(2)} = (0,b)$ for a plane of $\{001\}$ type in a simple cubic lattice,  and
$\mathbf{b}^{(1)} = \left(1,0\right)b$, $\mathbf{b}^{(2)} = \left(\frac{1}{2},\frac{\sqrt{3}}{2}\right)b$, $\mathbf{b}^{(3)} = \left(\frac{1}{2},-\frac{\sqrt{3}}{2}\right)b$ (in $xy$ coordinates) for a plane of $\{111\}$ type in a face-centered cubic (fcc) lattice.

\begin{figure}[htbp]
\centering
    \includegraphics[width=.35\linewidth]{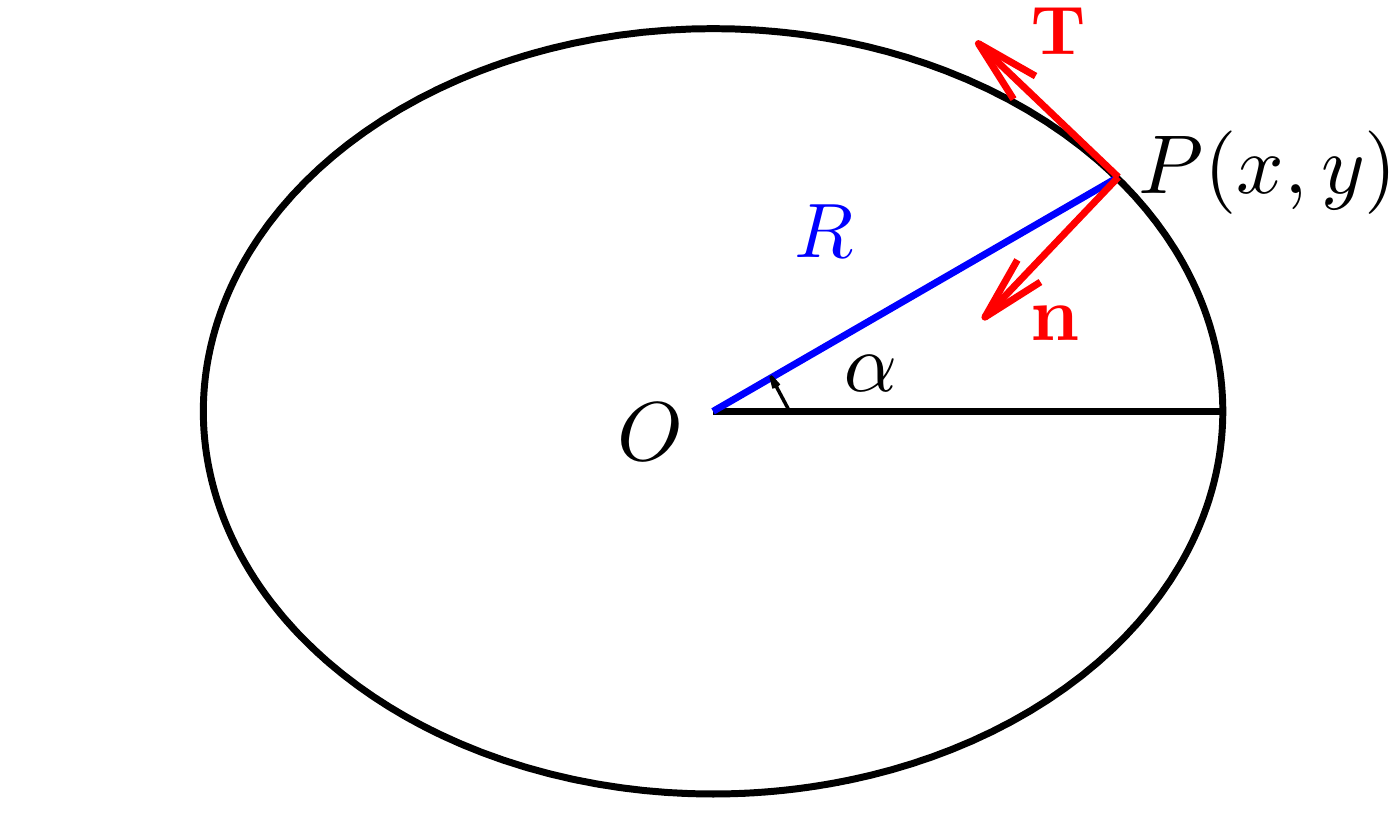}
    \caption{A two dimensional grain boundary $\Gamma$ with geometric center $O$. At a point $P(x,y)$ on the grain boundary, the normal and tangent directions on the grain boundary are $\mathbf n$ and $\mathbf T$, respectively,  $\alpha$ is the polar angle and $R$ is the radius. }
    \label{fig:geometry}
\end{figure}

We set the geometric center  of the grain boundary $\Gamma$ to be the origin $O$. The grain boundary $\Gamma$  can be parameterized  by its arclength $s$ (or sometimes by the polar angle $\alpha$).
 A point $P=(x,y)$ on the grain boundary has radius $R$, normal direction $\mathbf n$ and tangent direction $\mathbf T$, see Fig.~\ref{fig:geometry}. Using the arclength parameter $s$, we have
 $\mathbf T=\left(\frac{dx}{ds}, \frac{dy}{ds}\right)$, $\mathbf n=\left(-\frac{dy}{ds}, \frac{dx}{ds}\right)$, and $\kappa=\frac{dx}{ds}\frac{d^2 y}{ds^2} -\frac{dy}{ds}\frac{d^2 x}{ds^2}$,
where $\kappa$ is the curvature of the grain boundary.

\section{Review of the continuum model of grain boundary motion in Ref.~\cite{ZhangXiang2018157}}\label{sec:review}

In this section, we review our continuum model for grain boundary coupling and sliding motions with long-range dislocation interaction  presented in Ref.~\cite{ZhangXiang2018157}. The continuum formulation consists of the motion of the grain boundary and the evolution of the dislocation structure on the grain boundary. The continuum model is able to describe the increase of energy density due to the coupling motion and the decrease of energy density due to the sliding motion. The continuum model is derived from the discrete dislocation dynamics model for the motion and reaction of the constituent dislocations of the grain boundary, and has been validated by comparisons with simulations of atomistic models and discrete dislocation dynamics. The driving force comes from a total energy that consists of the energy of the long-range elastic interaction between the constituent dislocations of the grain boundary, the grain boundary local energy (line energy of the constituent dislocations), and the energy due to other effects such as the applied stress \cite{Zhu2014175}. For simplicity of description, in the review here, we only consider the driving forces  due to the long-range dislocation interaction and the local energy of the grain boundary. (See Eqs.~(12) and (13) in Ref.~\cite{ZhangXiang2018157} for the full model and the remark on including this driving force in the new formulation in Sec.~\ref{subsec:constraint}.)

With the problem settings  described in the previous section,  the motion of the grain boundary and the evolution of the dislocation structure on it are given by~\cite{ZhangXiang2018157}:

\vspace{0.05in}
\underline{Formulation with long-range dislocation interaction ({\bf F0})}
\begin{flalign}
v_{\rm n}=&M_{\rm d}\sum_{j=1}^J{\textstyle \frac{|\rho^{(j)}|}{\sum_{k=1}^J|\rho^{(k)}|}}\left({\mathbf f}_{\rm long}^{(j)}+{\mathbf f}_{\rm local}^{(j)}\right)\cdot \mathbf n,\label{eqn:vn}\\
\frac{\partial \rho^{(j)}}{\partial t}=&-\frac{\partial}{\partial s}\left(  M_{\rm d} ({\mathbf f}_{\rm long}^{(j)}\cdot \mathbf T)\rho^{(j)}\right)
-M_{\rm r}\frac{\partial \gamma}{\partial \rho^{(j)}}+v_{\rm n}\kappa \rho^{(j)},  \ \ j=1,2,\cdots,J, \label{eqn:vp}
\end{flalign}

Equation \eqref{eqn:vn} gives  the normal velocity of the grain boundary $v_{\rm n}$, which comes from the motion of the constituent dislocations in the normal direction of the grain boundary.
In this equation, ${\mathbf f}_{\rm long}^{(j)}$ and ${\mathbf f}_{\rm local}^{(j)}$
are the continuum long-range dislocation interaction force and the continuum local force,
respectively, on a dislocation with Burgers vector  $\mathbf{b}^{(j)}$, and $M_{\rm d}$ is the mobility of the dislocations
\footnote{As discussed in Ref.~\cite{ZhangXiang2018157},   available  atomistic simulations \cite{Cahn20021,Wu2012407,Voorhees2016264,Voorhees2017} were performed at high temperatures to examine the coupling motion of curved grain boundaries, which is purely geometric \cite{Cahn20044887}. At these temperatures, the dislocation climb mobility is comparable with that of dislocation glide. Following these simulations,
it is assumed in the continuum model that the dislocation  mobility $M_{\rm d}$ is isotropic  to fully examine the purely geometric coupling motion of low angle grain boundaries.}.
The velocity of a point on the grain boundary is the weighted average of the velocities of dislocations on the grain boundary with different Burgers vectors.

Equation \eqref{eqn:vp} describes the evolution of the dislocation structure on the grain boundary.  This lateral motion of the constituent dislocations is driven by the continuum long-range interaction force and dislocation reaction.
The first term on the right-hand side of Eq.~\eqref{eqn:vp}
describes the motion of dislocations along the grain boundary following conservation law
driven by the continuum long-range elastic force, and its
 importance  is to maintain a stable dislocation structure. Recall that the equilibrium dislocation structure, which is reached if and only if the long-range elastic fields cancel out, is governed by the Frank's formula \cite{Frank,Bilby,Zhu2014175}.
The second term  on the right-hand side of  Eq.~\eqref{eqn:vp} comes from the driving force due to variation of the grain boundary energy density  $\gamma$ (when the long-range elastic interaction vanishes)  with respect to the change of dislocation density $\rho^{(j)}$ on the fixed grain boundary, and $M_{\rm r}$ is the  mobility associated with this driving force.

Note that the first term on the right-hand side of Eq.~\eqref{eqn:vp} gives a motion by conservation law of the constituent dislocations, leading to the coupling motion of the grain boundary; and in the process of dislocation reaction described by the second term on the right-hand side of Eq.~\eqref{eqn:vp}, the constituent dislocations are not conserved, resulting in the sliding motion of the grain boundary.
The third term on the right-hand side of Eq.~\eqref{eqn:vp} is the change of dislocation density  due to the change of the arclength parameter $s$ as the grain boundary evolves, where $\kappa$ is the curvature of the grain boundary. More detailed discussion can be found in Ref.~\cite{ZhangXiang2018157}.

 In these evolution equations,  ${\mathbf f}_{\rm long}^{(j)}$ is the force that comes from the long-range interaction energy $E_{\rm long}$ (Eq.~(19) in Ref.~\cite{Zhu2014175}).
 This continuum long-range interaction force on a dislocation  of the $j$-th dislocation arrays located at the point $(x,y)$ is
 \begin{eqnarray}\label{eqn:flong0}
{\mathbf f}_{\rm long}^{(j)}(x,y)=\sum_{k=1}^J {\textstyle \frac{\rho^{(j)}(x,y)}{|\rho^{(j)}(x,y)|}}
\int_\Gamma\mathbf{f}^{(j,k)}(x,y;x_1,y_1)\rho^{(k)}(x_1,y_1)ds,
\end{eqnarray}
where $ds$ is the line element of the integral along the grain boundary $\Gamma$, the point $(x_1,y_1)$ varies along $\Gamma$ in the integral, and $\mathbf{f}^{(j,k)}(x,y;x_1,y_1)$ is the force acting on a dislocation  with Burgers vector $\mathbf b^{(j)}$ at the point $(x,y)$ generated by a dislocation with Burgers vector $\mathbf b^{(k)}$ at the point $(x_1,y_1)$, which is \cite{HL}
\begin{flalign}\label{eqn:flong00}
\mathbf{f}^{(j,k)}&=(f_1^{(j,k)}, f_2^{(j,k)}),
\end{flalign}
\begin{flalign}\label{eqn:flong01}
 f_1^{(j,k)}(x,y;x_1,y_1)
 =&  {\textstyle \frac{\mu}{2\pi(1-\nu)}\frac{1}{[(x-x_1)^2+(y-y_1)^2]^2}}
 \left(F_{1}+F_{2}+F_{3}+F_{4}\right),  \\
f_2^{(j,k)}(x,y;x_1,y_1)
 =& {\textstyle \frac{\mu}{2\pi(1-\nu)}\frac{1}{[(x-x_1)^2+(y-y_1)^2]^2}
\left(G_{1}+G_{2}+G_{3}+G_{4}\right),}   \label{eqn:flong1}
\end{flalign}
where
{\small
\begin{flalign*}
&F_1=  [(x-x_1)^3-(x-x_1)(y-y_1)^2]b_1^{(k)}b_1^{(j)}, \
F_2=[(x-x_1)^2(y-y_1) -(y-y_1)^3]b_2^{(k)}b_1^{(j)},\\
&F_3= [(x-x_1)^2(y-y_1)- (y-y_1)^3]b_1^{(k)}b_2^{(j)}, \
F_4= [(x-x_1)^3+3(x-x_1)(y-y_1)^2]b_2^{(k)} b_2^{(j)},\\
&G_1=  [3(x-x_1)^2(y-y_1)+(y-y_1)^3]b_1^{(k)} b_1^{(j)}, \
G_2=[-(x-x_1)^3+ (x-x_1)(y-y_1)^2]b_2^{(k)}b_1^{(j)}, \\
&G_3=[-(x-x_1)^3 +(x-x_1)(y-y_1)^2]b_1^{(k)}b_2^{(j)}, \
G_4=[-(x-x_1)^2(y-y_1) + (y-y_1)^3]b_2^{(k)}b_2^{(j)}.
\end{flalign*}}
Here $\mu$ is the shear modulus and $\nu$ is the Poisson ratio.
 The integral in Eq.~\eqref{eqn:flong0} is over all the grain boundaries if there are multiple grain boundaries in the system. Note that the long-range interaction force ${\mathbf f}_{\rm long}$ and the long-range interaction energy $E_{\rm long}$ vanish for an equilibrium grain boundary \cite{Sutton1995,HL,Zhu2014175}.

The continuum local force on a dislocation of the $j$-th dislocation arrays in these evolution equations is
\begin{eqnarray}\label{eqn:local_force}
{\mathbf f}_{\text{local}}^{(j)}\cdot\mathbf{n}={\textstyle \frac{\mu {b^{(j)}}^2}{4\pi(1-\nu)}}\kappa,
\end{eqnarray}
This local force comes from the grain boundary energy under equilibrium of its dislocation structure (which is the line energy of
the constituent dislocations; see $E_{\rm local}$ in Ref.~\cite{Zhu2014175,Zhang2017}), whose density is
\begin{flalign}
\gamma= \sum_{j=1}^J  {\textstyle \frac{\mu{b^{(j)}}^2}{4\pi(1-\nu)}} |\rho^{(j)}|
\log\! \frac{1}{r_g|\rho^{(j)}|},\label{eqn:gb_density}
\end{flalign}
where $r_g$ is a dislocation core parameter.
 Numerically,  $\log\! \frac{1}{|\rho^{(j)}|}$ can be regularized as $\log\! \frac{1}{\sqrt{{\rho^{(j)}}^2  +\epsilon}}$ with $\epsilon$ being a numerical cutoff parameter.

With the continuum model in Eqs.~\eqref{eqn:vn} and \eqref{eqn:vp} for the motion of the grain boundary and evolution of the dislocation structure on it, the misorientation angle $\theta$ between the two grains at any point on the grain boundary can be calculated  based on the Frank's formula  \cite{Frank,Bilby}, which is a condition for an equilibrium grain boundary dislocation structure  and is equivalent to the cancellation of the long-range elastic fields  \cite{Frank,Bilby,Zhu2014175}.
The misorientation angle $\theta$ is calculated by
 \begin{flalign}\label{eqn:theta_noncircular}
\theta
={\textstyle -\frac{1}{L}\int_0^{L}\sum_{j=1}^J \rho^{(j)}\left(\mathbf b^{(j)}\cdot{\mathbf n}\right)ds,}
\end{flalign}
where $L$ is the circumference of the grain boundary.

The local energy density $\gamma$ in Eq.~\eqref{eqn:gb_density} is a generalization of the classical Read-Shockley energy formula \cite{ReadShockley1950} that is widely used in the continuum models of grain  boundaries in the literature. In the Read-Shockley energy formula, grain boundary energy density depends on the misorientation angle $\theta$ as $\gamma=\gamma_0\theta(A-\ln\theta)$, where $E_0$ and $A$ are parameters. In our formulation in Eq.~\eqref{eqn:gb_density}, the energy density $\gamma$ depends on densities of the constituent dislocations,  which are related to the misorientation angle $\theta$ by Eq.~\eqref{eqn:theta_noncircular} (due to the long-range dislocation reaction and Frank's formula). In the new continuum formulations  to be presented in the next section, the dislocation densities are  related to the misorientation angle $\theta$ by the constraint of Frank's formula in Eq.~\eqref{eqn:frank} in the formulation (F1) and directly depends on the evolution of $\theta$ in Eqs.~\eqref{eqn:total_rho} and \eqref{eqn:delta_theta} in the formulation (F2).

{\bf Remark.} In our continuum model presented in Ref.~\cite{ZhangXiang2018157}, the dislocation density per unit polar angle $\varrho^{(j)}$, $j=1,2,\cdots,J$, are used as the variables in the continuum grain boundary dynamics model. The relationship between these two densities is $\varrho(\alpha)=\rho(\alpha)\frac{ds}{d\alpha}$, where $\alpha$ is the polar angle as shown in Fig.~\ref{fig:geometry}.
The advantage of using the dislocation density per angle $\varrho^{(j)}$ is that $\varrho^{(j)}$ does not change directly with the change of the arclength parameter $s$ as the grain boundary evolves.

Another alternative formulation with the same advantage of independence on the arclength parameter $s$
is to use the dislocation density potential functions $\eta^{(j)}$'s for the constituent dislocations \cite{Zhu2014175,ZhangXiang2018157}. A dislocation density potential function $\eta$ is a scalar function defined on the  grain boundary such that the constituent dislocations with the same Burgers vector are given by the integer-valued contour lines of $\eta$: {$\eta=i$, where $i$ is an integer}. Using the dislocation density potential function, the dislocation density per unit length $\rho$ and the dislocation density per polar angle $\varrho$ on the grain boundary are
$\rho(s)=\frac{d\eta}{ds}$ and $\varrho(\alpha)=\eta'(\alpha)$.
Using dislocation density potential functions
$\eta^{(j)}$, $j=1,2,\cdots,J$ to describe the distributions of the constituent dislocations with $J$ Burgers vectors,
 the evolution equation of $\eta^{(j)}$ that is equivalent to Eq.~\eqref{eqn:vp}  is:
$\frac{\partial \eta^{(j)}}{\partial t}=- M_{\rm d}\left( {\mathbf f}_{\rm long}^{(j)}\cdot \mathbf T\right)\frac{\partial \eta^{(j)}}{\partial s}
-\int_0^sM_{\rm r}\frac{\partial \gamma}{\partial \rho^{(j)}}(u)du$,  $j=1,2,\cdots,J$.
This formulation has also been used in the numerical simulations in Ref.~\cite{ZhangXiang2018157}.

\section{New continuum models}\label{sec:new_model}

In this section, we present new continuum models for the coupling and sliding motions of grain boundaries that addresses the limitation in the calculation of the long-range elastic interaction in our previously developed continuum model \cite{ZhangXiang2018157} summarized in formulation (F0) of Eqs.~\eqref{eqn:vn} and \eqref{eqn:vp}. In the first new model, the long-range elastic interaction force is replaced by the constraint of Frank's formula which consists only of local equations (formulation (F1) in Sec.~\ref{subsec:constraint}). Based on this constraint evolution and numerical implementation using the projection method~\cite{Chong2012}, we further derived a continuum dynamics model in which explicit velocity formulas of the grain boundary and constituent dislocation dynamics are given without any constraints (formulation (F2) in Sec.~\ref{subsec:projection}). Finally, we discuss the formulation (F2) for the pure coupling motion, effect of the sliding motion, and relationship with the classical motion by curvature model, in terms of the mechanisms of dislocation motion and reaction incorporated in our continuum models (Sec.~\ref{sec:modeldiscussion}).

\subsection{Continuum model in the form of constrained evolution}\label{subsec:constraint}

A key ingredient of the continuum formulation (F0) in Eqs.~\eqref{eqn:vn} and \eqref{eqn:vp} is the long-range elastic interaction between the constituent dislocations of the grain boundaries.
In fact, the long-range dislocation interaction  is so strong that an equilibrium state described by the Frank's formula~\cite{Frank,Bilby,Zhu2014175} is quickly reached during the evolution of the grain boundaries.
This behavior has been observed in simulations using the continuum model (F0) and the discrete dislocation dynamics model (from which the continuum model is derived) reported in Ref.~\cite{ZhangXiang2018157}.

 Now we make an assumption for simplification that the Frank's formula always holds during the evolution of the grain boundaries, which will not lead to significant change in grain boundary motion based on the above observations.
This assumption is implemented by replacing the long-range elastic interaction  by the constraint of Frank's formula in the continuum dynamics model. In the new formulation, since the equations in the Frank's formula are local, the time-consuming calculation of the long-range dislocation interaction is avoided.
This new continuum formulation is:

\vspace{0.05in}
\underline{Formulation of constrained evolution ({\bf F1})}
\begin{flalign}
v_{\rm n}=&M_{\rm d}C_\rho\kappa,
\label{eqn:vn_pj}\\
&{\rm where} \ C_\rho={\textstyle \frac{\mu}{4\pi(1-\nu)}\frac{\sum_{k=1}^J{b^{(j)}}^2|\rho^{(j)}|}{\sum_{k=1}^J|\rho^{(k)}|}=\frac{\mu b^2}{4\pi(1-\nu)}} \ {\rm if \ all} \ b^{(j)}=b,  \\
\frac{\partial \rho^{(j)}}{\partial t}=&-M_{\rm r}\frac{\partial \gamma}{\partial \rho^{(j)}}+v_{\rm n}\kappa\rho^{(j)}, \hspace{2ex} j=1,2,\cdots,J,
\vspace{1ex}
\label{eqn:vp_pj}\\
\text{subject to }\ \mathbf h=&\theta\mathbf n+\sum_{j=1}^J \rho^{(j)}\mathbf b^{(j)}=0.  \label{eqn:frank}
\end{flalign}
  Here the constraint in Eq.~\eqref{eqn:frank} is the Frank's formula.   Recall that the Frank's formula provides a link between the macroscopic degrees of freedom of the grain boundary and the microscopic dislocation structure, and is equivalent to cancellation of the long-range elastic fields  generated by the constituent dislocations of the grain boundary \cite{Frank,Bilby,Zhu2014175}.


In fact, after removing the long-range forces $\{{\mathbf f}_{\rm long}^{(j)}\}$ and keeping only the local forces $\{{\mathbf f}_{\rm local}^{(j)}\}$ in Eq.~\eqref{eqn:vn} of the continuum model (F0),  Eq.~\eqref{eqn:vn} becomes $v_{\rm n}=M_{\rm d}\sum_{j=1}^J\frac{|\rho^{(j)}|}{\sum_{k=1}^J|\rho^{(k)}|}{\mathbf f}_{\rm local}^{(j)}\cdot \mathbf n$. Using Eq.~\eqref{eqn:local_force}, 
we have Eq.~\eqref{eqn:vn_pj}.
The evolution equations \eqref{eqn:vn_pj} and  \eqref{eqn:vp_pj} can be obtained by variations of the grain boundary energy
$E= \int_\Gamma \gamma ds$, see Appendix B in Ref.~\cite{ZhangXiang2018157}.

{\bf Remark.} In addition to the terms in the formulation (F0), the continuum model with long-range dislocation interaction proposed in Ref.~\cite{ZhangXiang2018157} (Eqs.~(12) and (13) there) also includes the driving force of applied stress. To include this driving force in the constrained evolution formulation (F1), there will be one more term in the normal velocity $v_n$ in Eq.~\eqref{eqn:vn_pj}:
$M_{\rm d}\sum_{j=1}^J{\textstyle \frac{|\rho^{(j)}|}{\sum_{k=1}^J|\rho^{(k)}|}}{\mathbf f}_{\rm app}^{(j)} \cdot \mathbf n$, and one more term in the evolution of dislocation density in Eq.~\eqref{eqn:vp_pj}: $-\frac{\partial}{\partial s} \left( M_{\rm d}\rho^{(j)} \ {\mathbf f}_{\rm app}^{(j)}\cdot\mathbf T\right)$.
Here ${\mathbf f}_{\rm app}^{(j)}$ is the Peach-Koehler force \cite{HL} on a dislocation with Burgers vector $\mathbf b^{(j)}$ due to the applied stress $\pmb\sigma_{\rm app}$:
${\mathbf f}_{\rm app}^{(j)}=(\pmb\sigma_{\rm app}\cdot \mathbf{b}^{(j)})\times{\mathbf t}^{(j)}$,
with ${\mathbf t}^{(j)}$ being the line direction of the dislocation: ${\mathbf t}^{(j)}=\rho^{(j)}/|\rho^{(j)}|\hat{\mathbf z}$ (where $\hat{\mathbf z}$ is the unit vector in the $+z$ direction).

\subsection{Formulation based on projection method}\label{subsec:projection}
The constrained evolution model (F1) in Eqs.~\eqref{eqn:vn_pj}--\eqref{eqn:frank}  can be implemented using the projection method. In the projection method, the grain boundary and dislocation densities are evolved first without the constraint, and then the results are projected into a state that satisfies the constraint. We are able to find solution formula of this two-step projection procedure. This leads to the following formulation without constraint:

\vspace{0.05in}
\underline{Formulation of unconstrained evolution based on projection ({\bf F2})}
 \begin{flalign}
\mathbf v = &-\frac{1}{\theta}\frac{d\theta}{d t}(x,y)+\frac{1}{\theta }\left(- \sum_{j=1}^J s_\eta^{(j)}b_2^{(j)}, \ \sum_{j=1}^J s_\eta^{(j)}b_1^{(j)}\right),\label{eqn:total_v}\\
&{\rm where} \  \frac{d s_\eta^{(j)}}{ds}=s_\rho^{(j)}= -M_{\rm r}\frac{\partial \gamma}{\partial \rho^{(j)}}, \ j=1,2,\cdots,J,  \label{eqn:rho_rate} \vspace{1ex}\\
\frac{\partial \rho^{(j)}}{\partial t}
=&\frac{\rho^{(j)}}{\theta}\frac{d\theta}{dt}- \frac{\rho^{(j)}}{\theta}\left(\frac{dy}{ds}\sum_{j=1}^J s_\rho^{(j)}b_1^{(j)}-\frac{dx}{ds}\sum_{j=1}^J s_\rho^{(j)}b_2^{(j)}\right)+s_\rho^{(j)}, \label{eqn:total_rho} \\
\frac{d\theta}{d t} =& \frac{\theta}{2A}\int_\Gamma
M_{\rm d}C_\rho\kappa
ds -\frac{1}{2A}\int_\Gamma  \left(\frac{dx}{ds} \sum_{j=1}^J s_\eta^{(j)}b_1^{(j)}+\frac{dy}{ds}\sum_{j=1}^J s_\eta^{(j)}b_2^{(j)}\right)ds, \label{eqn:delta_theta}\\
&{\rm where}\ A={\textstyle \frac{1}{2}\int_L(x\frac{dy}{ds}-y\frac{dx}{ds})ds}. \label{eqn:areaA}
\end{flalign}
Here $\mathbf v$ is the velocity at point $(x,y)$  on the grain boundary, $\frac{\partial \rho^{(j)}}{\partial t}$ is the evolution of the density of the $j$-th dislocation array (with the $j$-th Burgers vector) on the grain boundary, $s_\rho^{(j)}$ is the rate of change of dislocation density $\rho^{(j)}$ due to dislocation reaction, $s_\eta^{(j)}(s)=\int_0^s s_\rho^{(j)} ds+s_\eta^{(j)}(0)$ and $s_\eta^{(j)}(0)$ is chosen such that $\int_\Gamma s_\eta^{(j)} ds=0$, $\frac{d\theta}{d t}$ is the evolution of the misorientation angle, and $A$ is the area of the enclosed grain when the grain boundary is closed. Note that in this formulation, $\mathbf v$ is the velocity vector, which is not necessarily in the normal direction of the grain boundary. (In fact, $s_\eta^{(j)}$ is the rate of change due to dislocation reaction in the dislocation density potential function $\eta^{(j)}$ \cite{Zhu2014175,ZhangXiang2018157}, see the review at the end of Sec.~\ref{sec:review}.)

 For simplicity in formulation, we focus on the case with $b^{(j)}=b$, $j=1,2,\cdots,J$, thus we have $C_\rho=\frac{\mu b^2}{4\pi (1-\nu)}$ and the first contribution to $\frac{d\theta}{d t}$ is $\frac{\theta}{2A}\int_\Gamma M_{\rm d}C_\rho\kappa ds=\frac{M_{\rm d}\mu b^2\theta}{4(1-\nu)A}$ (using $\int_\Gamma \kappa ds=2\pi$). We also assume that the dislocation density change rate due to reaction satisfies $\int_\Gamma s_\rho^{(j)}=0$, $j=1,2,\cdots,J$, e.g. due to some symmetry of the grain boundary. The origin $O$ is set to be the geometric center of the initial grain boundary, i.e. $\int_\Gamma xds=\int_\Gamma y ds=0$ when $t=0$.

In formulation (F2), we have evolution of the misorientation angle $\theta$ instead of the constraint of Frank's formula in the formulation (F1), in addition to  evolution of the grain boundary and evolution of the constituent dislocations on the grain boundary. The evolution of the misorientation angle $\theta$ is given in Eq.~\eqref{eqn:delta_theta}, which includes the contribution from the motion of grain boundary by the curvature force (the first term) and that due to dislocation reaction (the second term).

 The formulation (F2) is obtained by projecting the evolution of the grain boundary under the curvature force in Eq.~\eqref{eqn:vn_pj} and the evolution of dislocation densities given by Eq.~\eqref{eqn:vp_pj} to a state that satisfies the constraint of Frank's formula in Eq.~\eqref{eqn:frank}.
This projection procedure is solved based on the assumption that  the average of the normal velocity obtained by the projection procedure is the same as the average normal velocity due to the curvature driving force without constraint.  This assumption is equivalent to that the area of the inner grain is unchanged during the projection.
This assumption will be validated by comparisons with results of the full continuum model (F0) and discrete dislocation dynamics simulation, see Sec.~\ref{sec:numerical}. This assumption is also consistent with the theory of Taylor and Cahn based on mass transfer by surface diffusion \cite{Taylor2007493}. The meaning of this assumption is that the local misorientation angle $\theta$ calculated by Frank's formula after direct evolution with the curvature force and dislocation reaction is unified by the long-range force between dislocations (see formulation (F0)) with the average normal velocity unchanged.  The detailed derivation of the formulation (F2) from the formulation of constrained evolution (F1) is given below.

\subsubsection*{Derivation of the continuum formulation (F2)}\label{section:GB velocity}
 Assume that some time $t$, the Frank's formula in Eq.~\eqref{eqn:frank} is satisfied. After evolution of a small time step $\delta t$, when the Frank's formula is still satisfied, we have the following equations up to linear order of small changes:
\begin{equation}\label{eqn:delta_frank1}
\delta\theta\cdot\mathbf n+\theta\cdot\delta\mathbf n+\sum_{j=1}^J \delta\rho^{(j)}\mathbf b^{(j)}=0.
\end{equation}

 Suppose that at time $t$, a constituent dislocation of the grain boundary has velocity  $\mathbf v = v_{\rm n}\mathbf n+v_{\rm T}\mathbf T$, where $v_{\rm n}$ is the velocity in the normal direction of the grain boundary which is also the normal velocity of the grain boundary, and $v_{\rm T}$ is the velocity of the dislocation in the tangential direction of the grain boundary.
After the small time step $\delta t$, the point $(x,y)$ on the grain boundary and the dislocation structure on that point will move to a new position $(\tilde x,\tilde y)$. We have
$(\delta x, \delta y)=(\tilde x,\tilde y) - (x,y) =(v_{\rm n}\mathbf n+v_{\rm T}\mathbf T)\delta t$.

Using the arclength parameter $s$ of the grain boundary at time $t$, the unit normal vector of the grain boundary at time $t+\delta t$ is $\tilde {\mathbf n}=\frac{1}{\sqrt{\left(\frac{d \tilde x}{ds}\right)^2+\left(\frac{d \tilde y}{ds}\right)^2}}\left(-\frac{d \tilde y}{ds}, \frac{d \tilde x}{ds}\right)$. It can be calculated that
\begin{equation}\label{eqn:delta_n1}
\delta \mathbf n={\textstyle -\left(\frac{d v_{\rm n}}{ds}+\kappa v_{\rm T}\right)\mathbf T\delta t.}
\end{equation}
The arclength $\tilde s$ of the grain boundary at time $t+\delta t$ satisfies $\frac{d\tilde s}{ds}=\sqrt{\left(\frac{d \tilde x}{ds}\right)^2+\left(\frac{d \tilde y}{ds}\right)^2}=1+\left(-\kappa v_{\rm n}+\frac{dv_{\rm T}}{ds}\right)\delta t$ up to linear terms of $\delta t$. This gives
\begin{equation}\label{eqn:delta_s}
\delta(ds)={\textstyle \left(-\kappa v_{\rm n}+\frac{dv_{\rm T}}{ds}\right)\delta t \cdot ds.}
\end{equation}
 The dislocation density $\tilde \rho^{(j)}$ at $t+\delta t$ satisfies $\tilde \rho^{(j)} = \rho^{(j)}/\frac{d\tilde s}{ds}+s_{\rho}^{(j)}\delta t$, where $s_{\rho}^{(j)}$ is rate of change of $\rho^{(j)}$ due to dislocation reaction.
We have
\begin{equation}\label{eqn:delta_rho1}
\delta \rho^{(j)}={\textstyle\left( \kappa v_{\rm n}\rho^{(j)}- \frac{d v_{\rm T}}{ds}\rho^{(j)}+s_{\rho}^{(j)}\right)\delta t.}
\end{equation}

Substituting Eqs.~\eqref{eqn:delta_n1} and \eqref{eqn:delta_rho1} into Eq.~\eqref{eqn:delta_frank1}, using the Frank's formula $\theta\mathbf n+\sum_{j=1}^J \rho^{(j)}\mathbf b^{(j)}=0$ at time $t$, and dividing both sides by $\delta t$, we have
$\frac{\delta\theta}{\delta t}\mathbf n-\theta\left(\frac{d v_{\rm n}}{ds}+\kappa v_{\rm T}\right)\mathbf T
-\theta\left( \kappa v_{\rm n}- \frac{d v_{\rm T}}{ds}\right)\mathbf n+\sum_{j=1}^J s_{\rho}^{(j)}\mathbf b^{(j)}=0$.
Using $\frac{d\mathbf T}{ds}=\kappa \mathbf n$ and $\frac{d\mathbf n}{ds}=-\kappa \mathbf T$, it can be written as
\begin{equation}
{\textstyle \frac{\delta\theta}{\delta t}\mathbf n-\theta\frac{d}{ds}\left(v_{\rm n}\mathbf T\right)+
\theta\frac{d}{ds}\left( v_{\rm T}\mathbf n\right)+\sum_{j=1}^J s_{\rho}^{(j)}\mathbf b^{(j)}=0.}
\end{equation}
Integrating this equation, and using $\mathbf T=(\frac{dx}{ds},\frac{dy}{ds})$ and $\mathbf n=(-\frac{dy}{ds},\frac{dx}{ds})$, we can solving for $v_{\rm n}$ and $v_{\rm T}$ as
\begin{flalign}
v_{\rm n}=&{\textstyle \frac{1}{\theta}\frac{\delta\theta}{\delta t}\left(x\frac{dy}{ds}-y\frac{dx}{ds}\right)+\frac{1}{\theta}\sum_{j=1}^J \int_0^s s_{\rho}^{(j)} ds \left(b^{(j)}_1\frac{dx}{ds}+b^{(j)}_2\frac{dy}{ds}\right)}\label{eqn:vn3}\\
&{\textstyle +\frac{1}{\theta}\left(C_1\frac{dx}{ds}+C_2\frac{dy}{ds}\right)}, \nonumber
\end{flalign}
\begin{flalign}
v_{\rm T}=&{\textstyle  -\frac{1}{\theta}\frac{\delta\theta}{\delta t}\left(x\frac{dx}{ds}+y\frac{dy}{ds}\right)+\frac{1}{\theta}\sum_{j=1}^J \int_0^s s_{\rho}^{(j)} ds \left(b^{(j)}_1\frac{dy}{ds}-b^{(j)}_2\frac{dx}{ds}\right)}\label{eqn:vt3}\\
&{\textstyle +\frac{1}{\theta}\left(C_1\frac{dy}{ds}-C_2\frac{dx}{ds}\right)},\nonumber
\end{flalign}
where $C_1$ and $C_2$ are integration constants independent of $s$. The velocity $\mathbf v = v_{\rm n}\mathbf n+v_{\rm T}\mathbf T$ is
\begin{equation}\label{eqn:v4}
\mathbf v={\textstyle -\frac{1}{\theta}\frac{\delta\theta}{\delta t}(x,y)+\frac{1}{\theta}\left(-\sum_{j=1}^J b^{(j)}_2\int_0^s s_{\rho}^{(j)} ds -C_2, \  \sum_{j=1}^Jb^{(j)}_1 \int_0^s s_{\rho}^{(j)} ds+C_1\right).}
\end{equation}

We solve the constrained evolution problem (F1) by projection method from $t$ to $t+\delta t$. In the first step, the grain boundary $\Gamma$ evolves to a  tentative state under the curvature driving force given in Eq.~\eqref{eqn:vn_pj} without the constraint. Note that both the continuum long-range force and the curvature force in the original formulation (F0) have averages $\mathbf 0$. This is because these continuum forces come from interactions between the constituent dislocations \cite{Zhu2014175}, and the summation of all the interaction forces between these dislocations themselves should vanish. The obtained velocity  $\mathbf v$ in Eq.~\eqref{eqn:v4} after projection, which is proportional to the driving forces between the constituent dislocations, should also have average $\mathbf 0$, i.e., $\int_\Gamma \mathbf v ds=(0,0)$. This determines the two constants $C_1$ and $C_2$.

We first consider the case without dislocation reaction. In this case the velocity in Eq.~\eqref{eqn:v4} is $\mathbf v= -\frac{1}{\theta}\frac{d\theta}{d t}(x,y)+\frac{1}{\theta}\left(-C_2, C_1\right)$. It can be calculated from Eqs.~\eqref{eqn:vn3} and \eqref{eqn:vt3} that $-\kappa v_{\rm n}+\frac{dv_{\rm T}}{ds}=-\frac{1}{\theta}\frac{d\theta }{dt}$. By Eq.~\eqref{eqn:delta_s}, we have
$\frac{d}{dt}\int_\Gamma x ds=\int_\Gamma \frac{dx}{dt}ds+\int_\Gamma x\left(-\frac{1}{\theta}\frac{d\theta }{dt}\right)ds=-\frac{2}{\theta}\frac{d\theta }{dt} \int_\Gamma x ds$. Since initially, the geometric center of the grain boundary $(\int_\Gamma xds, \int_\Gamma yds)=(0,0)$, we have $\int_\Gamma xds=0$ for all time $t$. Similarly, we have  $\int_\Gamma yds=0$ for all time $t$. This means that the geometric center of the grain boundary $(\int_\Gamma xds, \int_\Gamma yds)=(0,0)$ does not change during the evolution. Using the condition $\int_\Gamma \mathbf v ds=(0,0)$, we have $C_1=C_2=0$ for all time $t$. Next consider the case when there are dislocation reactions, i.e., some $s_\rho^{(j)}\neq 0$. In this case,
if we define $s_\eta^{(j)}(s)=\int_0^s s_\rho^{(j)} ds+s_\eta^{(j)}(0)$ and choose $s_\eta^{(j)}(0)$ such that $\int_\Gamma s_\eta^{(j)} ds=0$, we have $\int_\Gamma \mathbf v ds=(0,0)$, with $C_1=\sum_{j=1}^J s_\eta^{(j)}(0) b^{(j)}_1$ and $C_2=\sum_{j=1}^J s_\eta^{(j)}(0) b^{(j)}_2$.


For the changes of dislocation densities, using Eqs.~\eqref{eqn:delta_rho1},  \eqref{eqn:vn3} and \eqref{eqn:vt3}, we have
\begin{equation}
{\textstyle \frac{\delta \rho^{(j)}}{\delta t}
=\frac{\rho^{(j)}}{\theta}\frac{\delta\theta}{\delta t}- \frac{\rho^{(j)}}{\theta}\left(\frac{dy}{ds}\sum_{j=1}^J s_\rho^{(j)}b_1^{(j)}-\frac{dx}{ds}\sum_{j=1}^J s_\rho^{(j)}b_2^{(j)}\right)+s_\rho^{(j)}.} \label{eqn:total_rho4}
\end{equation}

As discussed above (in the paragraph before this derivation),
the average of the normal component of the velocity $\mathbf v$ in Eq.~\eqref{eqn:v4} obtained by projection should be the same as the average of the normal velocity $v_{\rm n}$ due to curvature force without constraint before the projection given in Eq.~\eqref{eqn:vn_pj}, which is $\int_\Gamma \mathbf v\cdot \mathbf n ds=\int_\Gamma M_{\rm d} C_\rho \kappa ds$. This determines $\frac{\delta \theta}{\delta t}$ as
\begin{equation}
{\textstyle \frac{\delta \theta}{\delta t} = \frac{\theta}{2A}\int_\Gamma
M_{\rm d}C_\rho\kappa
ds -\frac{1}{2A}\int_\Gamma  \left(\frac{dx}{ds} \sum_{j=1}^J s_\eta^{(j)}b_1^{(j)}+\frac{dy}{ds}\sum_{j=1}^J s_\eta^{(j)}b_2^{(j)}\right)ds,} \label{eqn:delta_theta4}
\end{equation}
where $A$ is given by Eq.~\eqref{eqn:areaA}.
Letting $\delta t\rightarrow 0$ in Eqs.~\eqref{eqn:v4}, \eqref{eqn:total_rho4}, \eqref{eqn:delta_theta4}, we have the equations in formulation (F2).

\subsection{Discussion on different kinds of motions}\label{sec:modeldiscussion}

\subsubsection{Pure coupling motion with conservation of dislocations}\label{sec:coupling}
 When there is no dislocation reaction, in the formulation of constrained evolution (F1), the grain boundary moves by the curvature force in Eq.~\eqref{eqn:vn_pj} subject to the Frank's formula in Eq.~\eqref{eqn:frank}, and the dislocation densities change as the length of the grain boundary changes following Eq.~\eqref{eqn:vp_pj} with $M_{\rm r}=0$ during the evolution.
This is a pure coupling motion, with the number of dislocations being conserved. The continuum formulation (F2) in this case is

\underline{Pure coupling motion of grain boundary (FC2)}
\begin{flalign}
&\mathbf v = -\frac{1}{\theta}\frac{d\theta}{d t}(x,y),
\label{eqn:total_v0}\\
&\frac{\partial \rho^{(j)}}{\partial t}
=\frac{\rho^{(j)}}{\theta}\frac{d\theta}{d t}\label{eqn:total_rho0},
\  j=1,2,\cdots,J,  \vspace{1ex}\\
&\frac{d\theta}{d t} =  \frac{M_{\rm d} C_\rho\pi\theta}{A}. \label{eqn:total_theta0}
\end{flalign}
Recall that $A$ is given by Eq.~\eqref{eqn:areaA} and is the area of the enclosed grain when the grain boundary is closed, and $C_\rho=\frac{\mu b^2}{4\pi(1-\nu)}$.

This  motion is in  the inward radial direction of the grain boundary, governed by the velocity in Eq.~\eqref{eqn:total_v0}. Such behavior of the pure coupling motion agrees with the theoretical analyses  in Refs.~\cite{Taylor2007493,ZhangXiang2018157}  and simulations using the full continuum model (F0) and discrete dislocation dynamics simulations as shown in Ref.~\cite{ZhangXiang2018157}.
As can be seen from Eq.~\eqref{eqn:total_theta0} that the misorientation angle $\theta$ increases during the evolution, which agrees with the feature observed in molecular dynamics/phase field crystal simulations \cite{Cahn20021,Trautt20122407,Wu2012407} and discrete dislocation dynamics simulations \cite{ZhangXiang2018157} and that predicted by theories \cite{Cahn20044887,Taylor2007493,ZhangXiang2018157}. This formula of evolution of misorientation angle $\theta$ applies to a grain boundary with any shape, which generalizes the formula for a circular grain by pure coupling motion that available in the literature \cite{Cahn20021,Cahn20044887,Trautt20122407,Wu2012407}. Note that this pure coupling motion of a grain boundary is completely different from the classical motion by curvature theories reviewed in the introduction section, in which the misorientation angle $\theta$ is fixed during the evolution and the grain boundary velocity is in its normal direction and proportional to its curvature.

Now we briefly discuss the coupling effect of this kind of grain boundary motion. The two grains have a translation relative to each other during the motion of the grain boundary, whose tangential translation velocity is \cite{Taylor2007493}
\begin{equation}\label{eqn:v_parallel}
v_\parallel={\textstyle R\frac{d\theta}{d t}\cos\lambda,}
\end{equation}
where $\lambda$ is the angle between the inward radial direction and the normal direction,
$\cos\lambda= (-\frac{x}{R},-\frac{y}{R})\cdot(-\frac{dy}{ds},\frac{dx}{ds})=\frac{x}{R}\frac{dy}{ds}-\frac{y}{R}\frac{dx}{ds}$
with $R=\sqrt{x^2+y^2}$, see Fig.~\ref{fig:geometry}.
 By Eq.~\eqref{eqn:total_v0}, the grain boundary normal velocity of this motion $v_{\rm n}=\mathbf v\cdot(-\frac{dy}{ds},\frac{dx}{ds})=\frac{R}{\theta}\frac{d\theta}{d t}\cos\lambda$. Thus we have the relation $v_\parallel=\theta v_{\rm n}$. This agrees with the Cahn-Taylor theory \cite{Cahn20044887} with coupling parameter $\beta=\theta$. Recall that the meaning of the coupling motion is that the normal motion of the grain boundary induces tangential translation of the two grains.

 Next, we discuss some quantitative properties of the coupling motion of grain boundary described by Eqs.\eqref{eqn:total_v0}--\eqref{eqn:total_theta0} of formulation (FC2).

 {\bf (i) Shape preserving motion.}
By Eq.~\eqref{eqn:total_v0}, the velocity of the pure coupling motion is in  the inward radial direction of the grain boundary. Using polar coordinates $(R,\alpha)$ in which the grain boundary is $R=R(\alpha)$ (recalling that the origin is the geometric center of the grain boundary), the velocity in Eq.~\eqref{eqn:total_v0} can be written as
 \begin{equation}\label{eqn:vr}
{\textstyle \frac{\partial R}{\partial t} = -\frac{1}{\theta}\frac{d\theta}{d t}R.}
 \end{equation}

 From this equation, the ratio $\frac{\partial R}{\partial t}/R = -\frac{1}{\theta}\frac{d\theta}{d t}$ is independent of $\alpha$ (location of the point on the grain boundary). This means that the shape of the grain boundary is preserved as it shrinks during the evolution. This shape preserving property of the pure coupling motion agrees with the theoretical analyses  in Refs.~\cite{Taylor2007493,ZhangXiang2018157}  and simulations using the full continuum model  (F0) and discrete dislocation dynamics simulations as shown in Ref.~\cite{ZhangXiang2018157}.

{\bf (ii) Relationships between geometric quantities and misorientation angle during evolution.} Integrating Eq.~\eqref{eqn:vr}, we have
\begin{equation}\label{eqn:R-theta}
R(\alpha,t)\theta(t)=R(\alpha,0)\theta(0).
\end{equation}
This relationship between $R$ and $\theta$ during the pure coupling motion again agrees with the theoretical analyses  in Refs.~\cite{Taylor2007493,ZhangXiang2018157}  and simulations using the full continuum model  (F0) and discrete dislocation dynamics simulations as shown in Ref.~\cite{ZhangXiang2018157}.

Using the formulas of the total length of the grain boundary $L=\int_\Gamma\sqrt{R(\alpha)^2+R'(\alpha)^2}d\alpha$ and the area of the inner grain $A=\int_\Gamma \frac{1}{2}R(\alpha)^2d\alpha$, we have
\begin{flalign}
L(t)\theta(t)=L(0)\theta(0), \ \
A(t)\theta(t)^2=A(0)\theta(0)^2.\label{eqn:area-theta}
\end{flalign}

{\bf (iii) Velocity in the coupling motion.}
By Eqs.~\eqref{eqn:total_theta0} and \eqref{eqn:vr}, we have
\begin{equation}
{\textstyle \frac{\partial R}{\partial t} = -\frac{M_{\rm d}C_\rho \pi R}{A}.}
 \end{equation}
When the grain  boundary is circular, using $A=\pi R^2$, its velocity is $v_{\rm n}=-\frac{\partial R}{\partial t} = \frac{M_{\rm d}C_\rho}{R}$, which is the same as the velocity of motion by curvature.
Same velocity formula has been used in the explanation of the mechanism of the coupling motion under the simple setting of a circular grain boundary and each constituent dislocation has  Burgers vector in the radial direction \cite{Cahn20021,Cahn20044887,Cahn20064953,Trautt20122407,Wu2012407}. Our formulation in Eq.~\eqref{eqn:total_theta0} or \eqref{eqn:vr} applies to a grain boundary with general shape and realistic settings of Burgers vectors.

 Note that although the normal velocity in the coupling motion is the same as that in the motion by curvature for a circular grain boundary, the energy dissipation is quite different, see the discussion below. The energy density $\gamma$ in Eq.~\eqref{eqn:gb_density} in the coupling motion is increasing as the misorientation angle $\theta$ and dislocation densities $\rho^{(j)}$ increase during the motion as described in Eqs.~\eqref{eqn:total_rho0} and \eqref{eqn:total_theta0}; whereas in the motion by curvature, the energy density and misorientation angle are fixed during the evolution.

 {\bf (iv) Solution of the coupling motion equations.} The formulation (FC2) of Eqs.\eqref{eqn:total_v0}--\eqref{eqn:total_theta0} for grain boundary coupling motion can be solved analytically. Using Eqs.~\eqref{eqn:total_theta0} and \eqref{eqn:area-theta}, we have
 $\frac{d\theta}{d t} =  \frac{M_{\rm d} C_\rho\pi\theta^3}{A(0)\theta(0)^2}$. The solution of $\theta$ is
 \begin{equation}\label{eqn:solution_theta}
 \theta(t)=\left(1-{\textstyle \frac{2M_{\rm d} C_\rho\pi}{A(0)}}t\right)^{-\frac{1}{2}}\theta(0).
 \end{equation}

 Using other equations in the formulation (FC2) and the equations obtained above, we can calculate that
 \begin{flalign}
 &R(\alpha,t)=\left(1-{\textstyle \frac{2M_{\rm d} C_\rho\pi}{A(0)}}t\right)^{\frac{1}{2}}R(\alpha,0),\\
&L(t)=\left(1-{\textstyle \frac{2M_{\rm d} C_\rho\pi}{A(0)}}t\right)^{\frac{1}{2}}L(0), \ \
 A(t)=\left(1-{\textstyle \frac{2M_{\rm d} C_\rho\pi}{A(0)}}t\right)A(0),
 \end{flalign}
 \begin{flalign}
 &\rho(\alpha,t)=\left(1-{\textstyle \frac{2M_{\rm d} C_\rho\pi}{A(0)}}t\right)^{-\frac{1}{2}}\rho(\alpha,0).\label{eqn:solution_rho}
 \end{flalign}
 The formulas hold until $R$ reduces to $0$.

{\bf (v) Energy dissipation  in coupling motion.} The total energy of the grain boundary is $E=\int_\Gamma \gamma ds$, where the energy density $\gamma$ is given by Eq.~\eqref{eqn:gb_density}. With the formulation (FC2) in Eqs.\eqref{eqn:total_v0}--\eqref{eqn:total_theta0} for the coupling motion, using the solution formulas in Eqs.~\eqref{eqn:solution_theta}--\eqref{eqn:solution_rho} and $ds_t=\left(1-{\textstyle \frac{2M_{\rm d} C_\rho\pi}{A(0)}}t\right)^{\frac{1}{2}}ds_0$ (where $s_t$ is the arclength parameter of the grain boundary at time $t$), it can be calculated that
\begin{equation}\label{eqn:e-dissipation}
E(t)=E(0)+E_1\log\left(1-{\textstyle \frac{2M_{\rm d} C_\rho\pi}{A(0)}}t\right),
\end{equation}
where $E_1=\frac{1}{2}\int_{\Gamma}\sum_{j=1}^J  \frac{\mu{b^{(j)}}^2}{4\pi(1-\nu)} |\rho^{(j)}|ds$ for the grain boundary $\Gamma$ at $t=0$.

For a circular grain boundary under the motion by curvature, its normal velocity is $v_{\rm n}=C_\kappa \kappa$ with constant grain boundary energy density $\gamma$. For such a motion, we have $\frac{dR}{dt}=-C_\kappa/R$, $R(t)=R(0)\sqrt{1-\frac{2C_\kappa}{R(0)^2}t}$, and the total energy $E(t)=E(0)\sqrt{1-\frac{2C_\kappa}{R(0)^2}t}$. From this example, we can see that the energy dissipation in the coupling motion is quite different from that in the motion by curvature.

\subsubsection{Sliding motion with dislocation reactions} The dislocation densities can be reduced by dislocation reaction to reduce the total energy.
The point-wise decreases of dislocation densities described by the rates $s_\rho^{(j)}$ in Eq.~\eqref{eqn:rho_rate} lead to point-wise decrease of the misorientation angle $\theta$, as can be seen from the Frank's formula in Eq.~\eqref{eqn:frank}.
The point-wise decrease of $\theta$ is unified by projection to a Frank's formula satisfying state (which is driven by the long-range force in formulation (F0)). This decrease of misorientation angle $\theta$ due to dislocation reaction is the second term in the evolution equation of $\theta$  in Eq.~\eqref{eqn:delta_theta} of formulation (F2).

Such projection procedure generates contributions  in both grain boundary velocity (which changes the shape of the grain boundary) and dislocation evolution along the grain boundary, in order to accommodate a unified misorietation angle $\theta$. The shape change accompanied by grain rotation due to sliding motion has been modeled by mass transfer via diffusion along the grain boundary in Ref.~\cite{Taylor2007493}; in our model, it is achieved by local motions of the constituent dislocations in both normal and tangential directions of the grain boundary.
In other available models in which evolution of misorientation angle is considered during the grain boundary motion, the misorientation angle always has uniform changes (grain rotation) to reduce the total energy by gradient flow \cite{Li1962,Shewmon1966,Harris19982623,Kobayashi2000,Upmanyu2006,Selim2016,Liu2019gb}. These models do not explicitly consider the grain boundary shape change accompanied by the rotation of grain. In our model, such shape changes are enabled by considering point-wise changes of dislocation densities instead of the uniform change of the misoritation angle in these models in the literature.

The grain boundary  sliding motion, the translation of the two grains along the grain boundary is not induced by normal motion of the grain boundary \cite{Li1962,Shewmon1966}. Using Eq.~\eqref{eqn:v_parallel}, and considering the normal component of the velocity in Eq.~\eqref{eqn:total_v} of formulation (F2), it can be calculated that the translation velocity between the two grains is
$v_\parallel=\theta v_{\rm n}+v_{\rm s}$, with $v_{\rm s}=-\frac{dx}{ds}\sum_{j=1}^J s_\eta^{(j)}b^{(j)}_1-\frac{dy}{ds} \sum_{j=1}^Js_\eta^{(j)}b^{(j)}_2$ being the tangential translation velocity due to sliding. This agrees with the Cahn-Taylor theory \cite{Cahn20044887} with detailed expression of $v_{\rm s}$.

\subsubsection{Relation with the motion by curvature}\label{subsec:motion_curvature}

The classical grain boundary dynamics models are based upon the motion by curvature (mean curvature in three dimensions) to reduce the total grain boundary energy, and the misorientation angle of the grain boundary is fixed \cite{Herring1951,Mullins1956,Sutton1995}. Here we show that by choosing special dislocation reaction rates, our formulation (F2) recovers this motion by curvature.

 First of all, it can be seen from Eq.~\eqref{eqn:delta_theta} of formulation (F2) that the misorientation angle $\theta$ does not change during the evolution if
 \begin{equation}\label{eqn:curvature_cond1}
  \theta M_{\rm d}C_\rho\kappa
 = {\textstyle \frac{dx}{ds} \sum_{j=1}^J s_\eta^{(j)}b_1^{(j)}+\frac{dy}{ds}\sum_{j=1}^J s_\eta^{(j)}b_2^{(j)}.}
\end{equation}
When Eq.~\eqref{eqn:curvature_cond1} holds, $\frac{d\theta}{dt}=0$, the normal component of the velocity in Eq.~\eqref{eqn:total_v} of formulation (F2) is
\begin{equation}
v_{\rm n}=\mathbf v \cdot \mathbf n={\textstyle \frac{1}{\theta }\left(\frac{dx}{ds} \sum_{j=1}^J s_\eta^{(j)}b_1^{(j)}+\frac{dy}{ds}\sum_{j=1}^J s_\eta^{(j)}b_2^{(j)}  \right)=M_{\rm d}C_\rho\kappa.}
\end{equation}
This is the motion by curvature.

Now we focus on the condition in Eq.~\eqref{eqn:curvature_cond1}. This condition holds if we choose $\sum_{j=1}^J s_\eta^{(j)}b_1^{(j)}=\theta M_{\rm d}C_\rho\kappa\frac{dx}{ds}$ and $\sum_{j=1}^J s_\eta^{(j)}b_2^{(j)}=\theta M_{\rm d}C_\rho\kappa\frac{dy}{ds}$. Recall that $s_\eta^{(j)}(s)=\int_0^s s_\rho^{(j)} ds+s_\eta^{(j)}(0)$. Taking derivative with respect to the arclength $s$ in these equations, they become
\begin{equation}\label{eqn:curvature_cond2}
{\textstyle \sum_{j=1}^J s_\rho^{(j)}b_1^{(j)}=\theta M_{\rm d}C_\rho\frac{d}{ds}\left(\kappa\frac{dx}{ds}\right), \ \ \
\sum_{j=1}^J s_\rho^{(j)}b_2^{(j)}=\theta M_{\rm d}C_\rho\frac{d}{ds}\left(\kappa\frac{dy}{ds}\right).}
\end{equation}
Therefore, if the dislocation density change rates $s_\rho^{(j)}$, $j=1,2,\cdots,J$, satisfy Eq.~\eqref{eqn:curvature_cond2}, our formulation (F2) reduces to the classical motion by curvature model. Note that the mechanism of dislocation reaction to maintain the motion by curvature has been discussed in Ref.~\cite{Rath2007}.

\section{Numerical simulations}\label{sec:numerical}
In this section, we validate our continuum formulation (F2) (Eqs.~\eqref{eqn:total_v}--\eqref{eqn:areaA}) for the motion of grain boundaries and evolution of their dislocation structures by numerical simulations and comparisons  with the results of the continuum model with long-range dislocation interaction (F0) proposed in Ref.~\cite{ZhangXiang2018157} and discrete dislocation dynamics simulations.

We start from a grain boundary $\Gamma$ with the equilibrium dislocation structure that satisfies the Frank's formula and has the lowest energy, which is calculated pointwisely using the method in Ref.~\cite{Zhang2017}.
We focus on the case that the $xy$ plane is a $(111)$ plane in fcc crystals, and the directions $[\bar{1}10], [\bar{1}\bar{1}2]$  are chosen to be the $+x$ and  $+y$ directions, respectively. There are $J=3$ possible Burgers vectors in this plane:
$\mathbf{b}^{(1)} = \left(1,0\right)b$, $\mathbf{b}^{(2)} = \left(\frac{1}{2},\frac{\sqrt{3}}{2}\right)b$, and $\mathbf{b}^{(3)} = \left(\frac{1}{2},-\frac{\sqrt{3}}{2}\right)b$.  The rotation axis of the grain  boundary is in the $+z$ direction, which is the $[111]$ direction in the fcc lattice.  We choose the Poisson's ratio $\nu=0.347$, which is the value of aluminum.

The grain boundary $\Gamma$ is discretized into $360$ grid points in the continuum models (F2) and (F0). Trapezoidal rule is used to calculate the integrals in the long-range force in the continuum formulation (F0). In discrete dislocation dynamics simulations in this two-dimensional setting, each dislocation is a point in the $xy$ plane, and the force acting on a dislocation  with Burgers vector $\mathbf b^{(j)}$ at the point $(x,y)$ generated by a dislocation with Burgers vector $\mathbf b^{(k)}$ at the point $(x_1,y_1)$ is given by Eqs.~\eqref{eqn:flong00}--\eqref{eqn:flong1}. These dislocations moves by the mobility law $\mathbf v=M_{\rm d}\mathbf f$.

\subsection{Pure coupling motion (with dislocation conservation)}\label{sec:coupling_numerical}
In this subsection, we focus on the evolution of grain boundaries with conservation of dislocations, i.e., there is no dislocation reaction:  $M_{\rm r}=0$, and the grain boundary motion is the pure coupling motion. In this case, the formulation (F2) is reduced to the formulation (FC2) of Eqs.~\eqref{eqn:total_v0}--\eqref{eqn:total_theta0}.

\subsubsection{Circular grain boundary}

\begin{figure}[htbp]
\centering
\includegraphics[width=.75\linewidth]{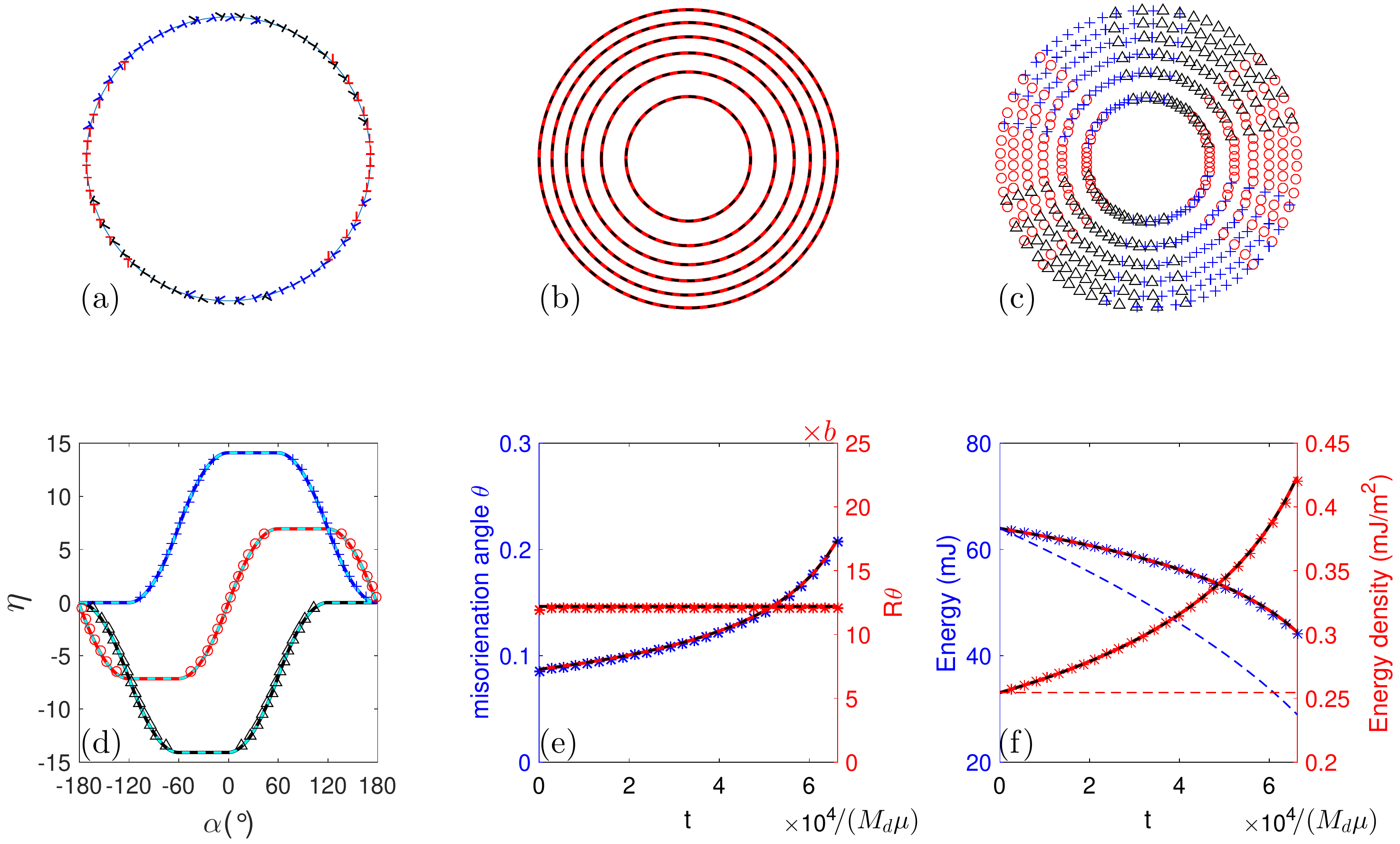}
     \caption{Grain boundary motion under dislocation conservation. The initial grain boundary is circular. (a) Equilibrium dislocation structure on the initial circular grain boundary. There are three arrays of dislocations with  Burgers vectors $\mathbf{b}^{(1)}$ (red), $\mathbf{b}^{(2)}$ (black), and $\mathbf{b}^{(3)}$ (blue), respectively.  (b) Motion of the grain boundary (shrinkage) by using our new continuum model (FC2) (red solid lines) and continuum model with long-range force (F0) (black dashed lines).  (c)  Motion of the grain boundary by using the discrete dislocation model. In (b) and (c),  the grain boundary is plotted at uniform time intervals starting with the outer most one. (d) Evolution of dislocation density potential functions of these three arrays of dislocations on the grain boundary (same colors as in (a)). The curves show the results of the new continuum model (FC2) (solid lines) and the continuum model with long-range force (F0)  (dashed lines); the dots show the results using the discrete dislocation model. (e) The curves show the evolution of the misorientation angle $\theta$, using the new continuum model (FC2) (red solid line), continuum model with long-range force (F0) (black dashed line), and discrete dislocation dynamics model (stars). The result using the solution formula  in Eq.~\eqref{eqn:solution_theta} is also plotted (blue solid line), and it is almost identical to the numerical result of model (FC2) (red solid line) and their difference cannot be seen from the figure. The straight lines in (e) show the results of $R\theta$ using the three models and the formula in Eq.~\eqref{eqn:R-theta} (with the same lines/symbols as the $\theta$-curves), and the vertical axis is on the right of the figure. (f) Evolutions of the total energy (curves with blue stars on them, vertical axis on the left) and average energy density (curves with red stars on them, vertical axis on the right), using the three models and the analytical formula in Eq.~\eqref{eqn:e-dissipation} with the same lines/symbols as in (e). Evolutions of the total energy and energy density  using the motion by curvature model (dashed lines, blue for total energy and red for energy density) are also shown in (e). }
    \label{circle_motion}
\end{figure}

We first consider the pure coupling motion of an initially circular grain boundary with radius $R_0=140b$ and misorientation angle  $\theta=5^\circ$. The calculated  equilibrium dislocation structure on this initial grain boundary is shown in Fig.~\ref{circle_motion}a.
 The distribution of these dislocations are shown in Fig.~\ref{circle_motion}d by the dislocation density potential functions $\eta^{(1)}$, $\eta^{(2)}$, and ${\eta^{(3)}}$ for dislocations with Burgers vectors $\mathbf{b}^{(1)}$, $\mathbf{b}^{(2)}$, and $\mathbf{b}^{(3)}$, respectively.
   Recall that the dislocation density potential functions are defined on the grain boundary such that the constituent dislocations are located at their integer-valued contour lines, and their derivatives with respect to the arclength parameter $s$ of the grain boundary are the dislocation densities  $\rho^{(j)}$, $j=1,2,3$.
In  Fig.~\ref{circle_motion}d, locations of dislocations in the discrete dislocation model are also shown using the dislocation density potential functions.

We perform simulations using our new formulation (FC2), the continuum formulation  with the long-range force (F0) \cite{ZhangXiang2018157}, and the discrete dislocation model.
Figure~\ref{circle_motion}b and \ref{circle_motion}c show the evolution  of this circular grain boundary obtained using these models. Excellent agreement can be seen from these results. During the evolution, the circular grain boundary shrinks with increasing rate and keeps the circular shape.

Evolution of the dislocation structure during the shrinking of this grain boundary by using these models are shown in Fig.~\ref{circle_motion}d, based on the dislocation density potential functions $\eta^{(1)}$, $\eta^{(2)}$, and ${\eta^{(3)}}$. Again, excellent agreement of the results using these three models can be seen. We can see from this figure that the dislocation density potential functions
maintain their initial profiles as functions of the polar angle during the evolution, meaning the conservation of dislocations.
This behavior of dislocation density potential functions also means that the locations of these dislocations with respect to the polar angle do not change, indicating
  that all the dislocations move in the inward radial direction of the grain boundary. This can also be seen from the result of the discrete dislocation model shown in Fig.~\ref{circle_motion}c.

Results of evolution of misorientation angle $\theta$ during the shrinkage of this circular grain boundary using the three models and  the solution formula  in Eq.~\eqref{eqn:solution_theta} are shown in Fig.~\ref{circle_motion}e. These four results  agree excellently. The misorientation angle $\theta$ is increasing during this coupling motion of grain boundary. In Fig.~\ref{circle_motion}e, we also show the evolution of the radius $R$ by examining the relation $R\theta=$constant in Eq.~\eqref{eqn:R-theta}. The results using these four methods agree excellently, which validates this relation.

Evolutions of the total energy and average energy density are shown in Fig.~\ref{circle_motion}f. It can be seen that the total energy is decreasing whereas the energy density is increasing under this pure coupling motion. The results obtained by the three models and the analytical formula in Eq.~\eqref{eqn:e-dissipation} agree excellently. In Fig.~\ref{circle_motion}f, we also show evolutions of the total energy and energy density  using the motion by curvature model (with isotropic energy density), in which the total energy of the initial grain boundary is the same as that in the coupling motion. It can be seen that the energy dissipation in the motion by curvature is faster than that in the pure coupling motion, and the energy density remains constant. These agree with the analytical discussion in Sec.~\ref{sec:coupling}.

These excellent agreements of the results with those of the discrete dislocation model and our previous continuum model (F0) with long-range elastic force validate  the new continuum model (F2) (reduced to (FC2) in this case), for the coupling motion of this circular grain boundary.

\subsubsection{Elliptic grain boundary}

\begin{figure}[htbp]
\centering
\includegraphics[width=.6\linewidth]{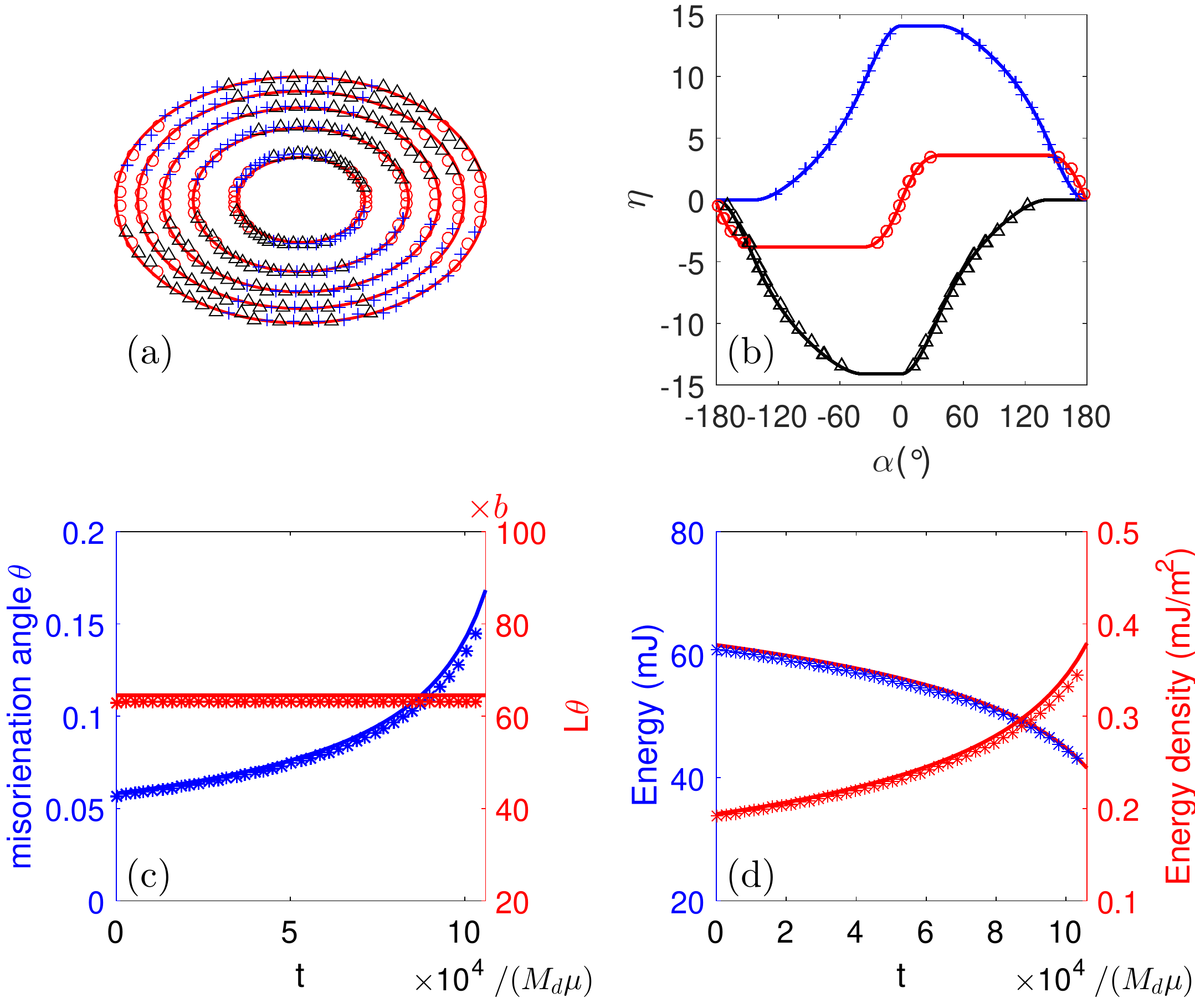}
     \caption{Grain boundary motion under dislocation conservation. The initial grain boundary is an ellipse.  (a) Motion of the grain boundary (shrinkage) by using our new continuum model (FC2) (red curves) and the discrete dislocation dynamics model (discrete symbols). The grain boundary is plotted at uniform time intervals starting with the outer most one. (b) Evolution of dislocation density potential functions $\eta^{(1)}$, $\eta^{(2)}$, and ${\eta^{(3)}}$,  which represent  dislocations on the grain boundary with  Burgers vectors $\mathbf{b}^{(1)}$ (red), $\mathbf{b}^{(2)}$ (black), and $\mathbf{b}^{(3)}$ (blue), respectively. The curves show the results of the new continuum model (FC2) and the discrete symbols show the results using the discrete dislocation model.
     (c) The curves show the evolution of the misorientation angle $\theta$, using the new continuum model (FC2) (blue solid line) and discrete dislocation dynamics model (stars). The result using the solution formula  in Eq.~\eqref{eqn:solution_theta} is also plotted (blue solid line), and it is almost identical to the numerical result of model (FC2) (blue dotted line) and their difference cannot be seen from the figure.
     The straight lines in (c) show the results of $L\theta$ using the two models and the formula in Eq.~\eqref{eqn:area-theta} (with red color and the same lines/symbols as the $\theta$-curves), and the vertical axis is on the right of the figure.
     (d) Evolutions of the total energy (curves with blue color, vertical axis on the left) and average energy density (curves with red color, vertical axis on the right), using the two models and the analytical formula in Eq.~\eqref{eqn:e-dissipation} with the same lines/symbols as in (c). }
    \label{ellipse_motion}
\end{figure}

Next, we consider the pure coupling motion of a grain boundary with initial shape of ellipse and misorientation angle $\theta=3.33^\circ$. The semi-minor axis of the ellipse is $140b$ and the ratio of the  major axis to the minor axis  is $1.5$, see the outer most curve in  Fig.~\ref{ellipse_motion}a. We perform simulations using the new continuum formulation (FC2) and the discrete dislocation dynamics model, and the simulation results are shown in  Fig.~\ref{ellipse_motion}. Excellent agreement can be seen from the results using these two models.

Evolution of this elliptic grain boundary is shown in Fig.~\ref{ellipse_motion}a. The elliptic grain boundary shrinks with increasing rate and its shape does not change during the evolution. Each dislocation moves in the inward radial direction of the grain boundary. These numerical results agree  with the theoretical predictions in Sec.~\ref{sec:coupling}. Here we can see that the shape-preserving property predicted by our new continuum formulation (FC2) is validated by discrete dislocation dynamics simulation, and it agrees with  the results using the continuum formulation  with  long-range force (F0) obtained in Ref.~\cite{ZhangXiang2018157} and the theoretical analysis  in Ref.~\cite{Taylor2007493} by using their model based on mass surface diffusion.
Evolution of dislocation densities are shown in Fig~\ref{ellipse_motion}b by dislocation density potential functions, which are unchanged during the evolution. This behavior is consistent with the shape-preserving motion of the grain boundary.

Fig.~\ref{ellipse_motion}c shows evolution of the misorientation angle $\theta$, which is increasing during the evolution.  Evolution of $\theta$ predicted by the solution formula  in Eq.~\eqref{eqn:solution_theta} is also plotted in Fig.~\ref{ellipse_motion}c, and the result is almost identical to the numerical result using the continuum model (FC2). In Fig.~\ref{ellipse_motion}c, we also show the evolution of the total length of the grain boundary $L$ by examining the relation $L\theta=$constant in Eq.~\eqref{eqn:area-theta}. The results using the two models and the analytical formula agree excellently, which validates this relation. Fig.~\ref{ellipse_motion}d shows that the total energy is decreasing whereas the average energy density is increasing during the evolution.  Results of the two models agree excellently with prediction of the analytical formula of energy dissipation in Eq.~\eqref{eqn:e-dissipation}, which is also plotted in  Fig.~\ref{ellipse_motion}d.

Evolution of this elliptic grain boundary is completely different from that by motion by curvature, in which the ellipse will evolve into a circle and the misorientation angle and energy density will remain constant during the evolution.

\subsubsection{Grain boundary with other shapes: hexagon and star polygon}

\begin{figure}[htbp]
\centering
\includegraphics[width=.5\linewidth]{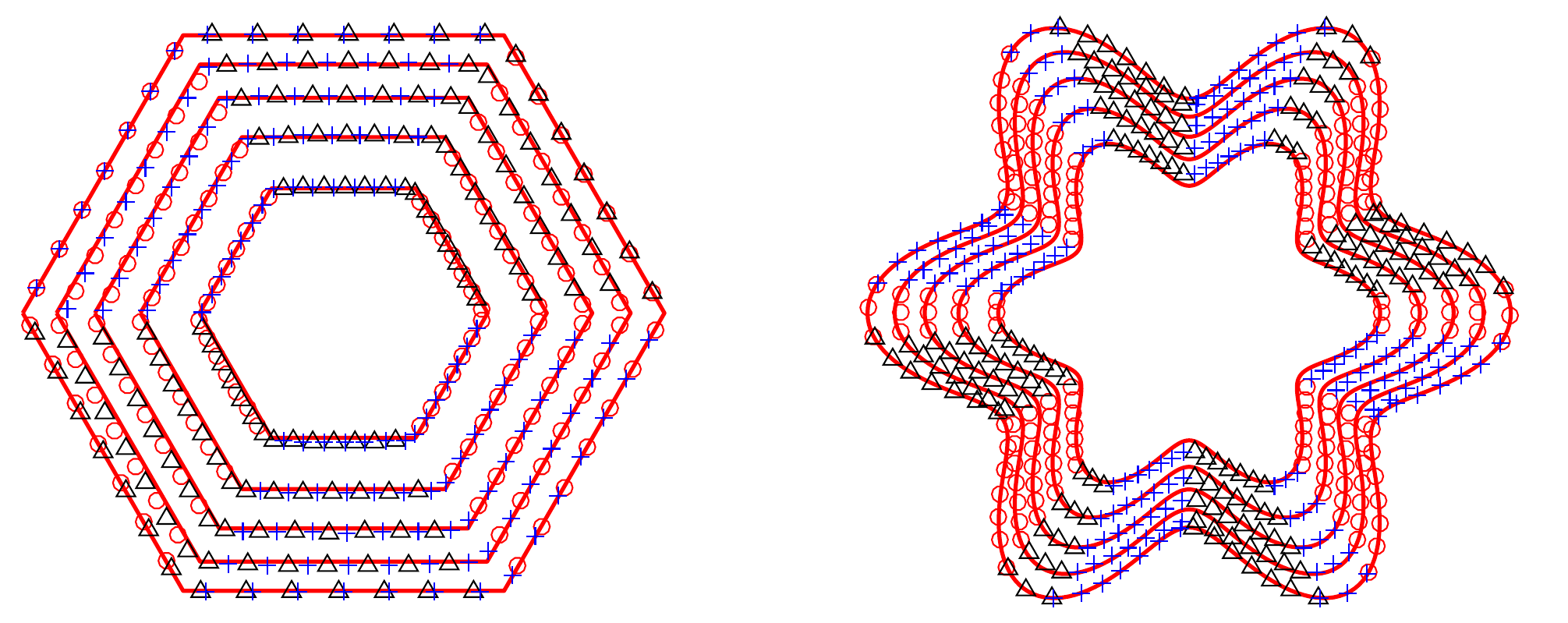}
     \caption{Motion of grain boundaries with shapes of hexagon and star polygon under the pure coupling motion,  by using our continuum model (FC2) (red curves) and the discrete dislocation dynamics model (discrete symbols, the three colors indicate dislocations with different Burgers vectors).}
    \label{fig:Hex_snow_motion}
\end{figure}

We also consider the coupling motion of grain boundaries with shapes of hexagon and star polygon. These shapes have the characters of sharp corners or concavities. The length of the longest diagonal line is $140b$ in both initial shapes, see the outer most curves in  Fig.~\ref{fig:Hex_snow_motion}.
We perform simulations using the new continuum formulation (FC2) and the discrete dislocation dynamics model, and the simulation results are shown in Fig.~\ref{fig:Hex_snow_motion}. Excellent agreement can be seen from these results using these two models.

The simulation results show that under the pure coupling motion, the hexagon and star polygon also shrink in a shape preserving way. The corners and concavities  on the grain boundary do not change during the evolution. All the constituent dislocations  move in the inward radial direction.
Again, the shape-preserving property predicted by our new continuum formulation (FC2) is validated by discrete dislocation dynamics simulations, and it agrees with  the results using the continuum formulation  with  long-range force (F0) obtained in Ref.~\cite{ZhangXiang2018157} and the theoretical analysis  in Ref.~\cite{Taylor2007493} by using their model based on mass surface diffusion. This shape-preserving, coupling motion of grain boundaries with hexagon and star polygon is completely different from the motion by curvature, in which the sharp corners and concavities will be smoothed out during the evolution and the grain boundary will evolve into a circle.

\subsection{Motion with grain boundary sliding (dislocation reaction)}

\begin{figure}[htbp]
\centering
\includegraphics[width=.7\linewidth]{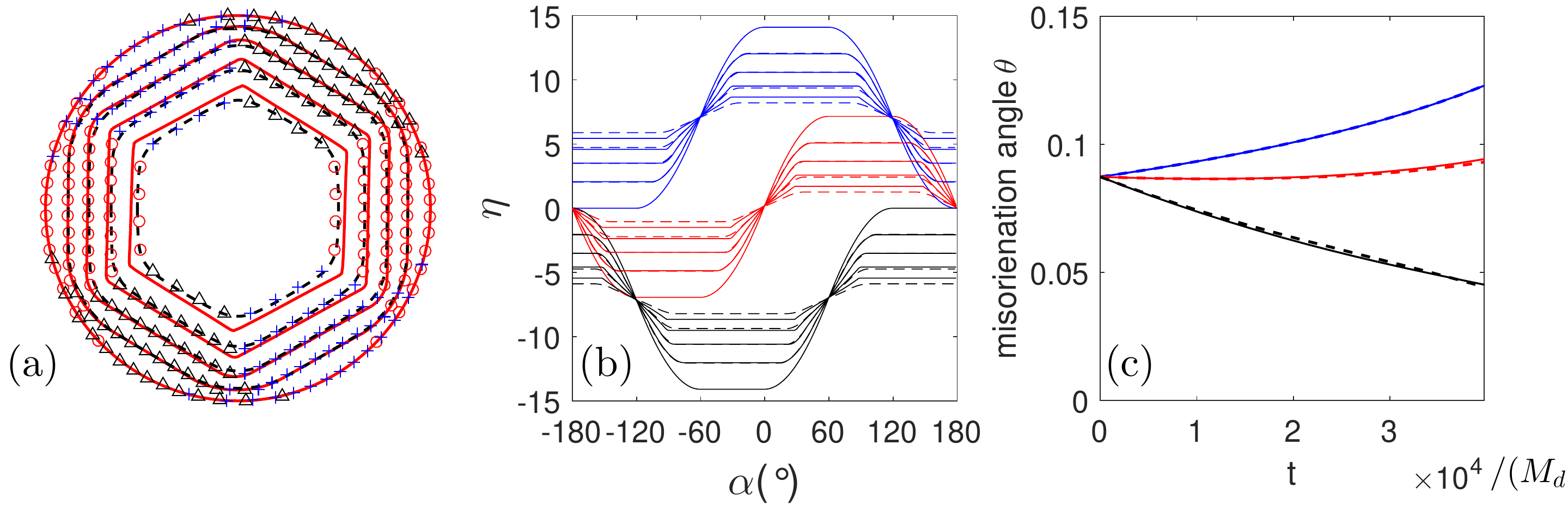}
     \caption{Motion of an initially circular grain boundary under both coupling and sliding (dislocation reaction mobility $M_{\rm r}\neq 0$) effects. (a) Evolution of the grain boundary by using the new continuum model (F2) (red lines), the continuum formulation  with  long-range force (F0) \cite{ZhangXiang2018157} (dashed lines), and the discrete dislocation dynamics model (discrete symbols). The three colors indicate dislocations with different Burgers vectors.  (b) Evolution of dislocation density potential functions by using  the new continuum model (FC2) (solid lines), and the continuum formulation  with  long-range force (F0) (dashed lines). The three colors indicate dislocations with different Burgers vectors.   In (a) and (b),  $M_{\rm r}b^3/ M_{\rm d}=4.74\times10^{-6}$. (c) Evolutions of the misorietation angle $\theta$ with different values of reaction mobility $M_{\rm r}b^3/ M_{\rm d}=0, 1.58\times10^{-6}, 4.74\times10^{-6}$ (from the top to the bottom).}
    \label{fig:circle_reaction}
\end{figure}

In this subsection, we perform simulations using our new continuum formulation (F2) for the case in which dislocation reaction is not negligible, i.e., $M_{\rm r}\neq 0$, and it generates sliding motion during the evolution of the grain boundary. Simulation results are compared with those of the continuum formulation  with the long-range force (F0) \cite{ZhangXiang2018157} and the discrete dislocation dynamics model. Simulation results from an initially circular grain boundary with radius $R_0=140b$ and misorientation angle  $\theta=5^\circ$ are shown in  Fig.~\ref{fig:circle_reaction}. The grain boundary is discretized into $180$ grid points in the simulations using the continuum models.  Excellent agreement can be seen from the results using these three models, which validates the new continuum formulation (F2).

Fig.~\ref{fig:circle_reaction}a and Fig.~\ref{fig:circle_reaction}b show evolutions of the grain boundary and the dislocation structure on it, with the mobility associated with dislocation reaction $M_{\rm r}b^3/ M_{\rm d}=4.74\times10^{-6}$.  It can be seen from Fig.~\ref{fig:circle_reaction}a that the initial circular grain boundary gradually changes to hexagonal shape as it shrinks. Each edge in this hexagon is a pure tilt boundary that consists of dislocations of only one Burgers vector. This behavior is consistent with the fact that the energy density of the grain boundary is anisotropic and the pure tilt boundary has the minimum energy of all tilt boundaries \cite{ReadShockley1950,Sutton1995,HL,Zhang2017}.
Results of evolution of the dislocation structure are shown in Fig.~\ref{fig:circle_reaction}b by evolution of the dislocation density potential functions. The amplitude of each dislocation density potential function $\eta^{(j)}$ is decreasing,  meaning that the dislocations react and the number of dislocations of each Burgers vector is reduced.
 Note that in the discrete dislocation dynamics simulation, we gradually remove two pairs of dislocations in each of the three arrays of dislocations with different Burgers vectors during the evolution according to the reaction rate in the continuum models. (In fact, the discrete dislocation dynamics model does not include the non-trivial  mechanisms of grain boundary dislocation reactions, e.g. by annihilation of dislocation pairs with opposite Burgers vectors located on opposite ends of the grain \cite{Cahn20021}, or by a chain of dissociation and association of the GB dislocations \cite{Trautt20122407}.)

Evolutions of the misorietation angle $\theta$ with different values of reaction mobility $M_{\rm r}$ are shown in Fig.~\ref{fig:circle_reaction}c. Evolution of $\theta$ in the new continuum formulation (F2) is given by Eq.~\eqref{eqn:delta_theta}. Recall that the first term in Eq.~\eqref{eqn:delta_theta} is due to the coupling motion of grain boundary associated with the conservation of dislocations, which increases $\theta$; and the second term in it is due to the sliding motion generated by dislocation reaction, which decreases $\theta$.
As can be seen from the simulation results in Fig.~\ref{fig:circle_reaction}c, when the dislocation reaction mobility $M_{\rm r}$ increases, meaning the sliding effect due to dislocation reaction is becoming stronger, the increase rate of $\theta$ is decreasing during the motion of the grain boundary, and when the sliding effect is strong enough, the misorientation angle $\theta$ is decreasing.

\section{Conclusions and discussion}
We have developed simplified formulations for the continuum model of two dimensional low angle grain boundary motion and  dislocation structure evolution on the grain boundary developed in Ref.~\cite{ZhangXiang2018157}, by replacing the long-range elastic interaction between dislocations in the continuum model  by a constraint of the Frank's formula.  The constrained evolution problem in our new continuum model is further solved by using the projection method, for which analytical formula has been found.  Effects of the coupling and sliding motions in our new continuum formulations and relationship with the classical motion by curvature model are discussed, in terms of the mechanisms of dislocation motion and reaction incorporated in our continuum models. The continuum model is validated by comparisons with discrete dislocation dynamics model and our early continuum model \cite{ZhangXiang2018157} in which the long-range dislocation interaction is explicitly calculated.
We remark that continuum models for the dynamics of high angle grain boundaries have also been developed \cite{Zhang2017119,Wei2019133} based on a disconnection model and atomistic simulations \cite{Thomas2017}.

\bibliographystyle{plain}
\bibliography{science_v2}

\begin{thebibliography}{10}

\bibitem{Admal2018}
N.~C. Admal, G.~Po, and J.~Marian.
\newblock A unified framework for polycrystal plasticity with grain boundary
  evolution.
\newblock {\em Int. J. Plasticity}, 106:1--30, 2018.

\bibitem{Ask2018}
A~Ask, S.~Forest, B.~Appolaire, K.~Ammar, and O.~U. Salman.
\newblock A cosserat crystal plasticity and phase field theory for grain
  boundary migration.
\newblock {\em J. Mech. Phys. Solids}, 115:167--194, 2018.

\bibitem{Bilby}
B.~A. Bilby.
\newblock In {\em Bristol conference report on defects in crystalline
  materials}, page 123. Physical Society, London, 1955.

\bibitem{Cahn20064953}
J.~W. Cahn, Y.~Mishin, and A.~Suzuki.
\newblock Coupling grain boundary motion to shear deformation.
\newblock {\em Acta Mater.}, 54:4953--4975, 2006.

\bibitem{Cahn20044887}
J.~W. Cahn and J.~E. Taylor.
\newblock A unified approach to motion of grain boundaries, relative tangential
  translation along grain boundaries, and grain rotation.
\newblock {\em Acta Mater.}, 52:4887--4898, 2004.

\bibitem{Chenlq1994}
L.~Q. Chen and W.~Yang.
\newblock Computer simulation of the domain dynamics of a quenched system with
  a large number of nonconserved order parameters: The grain-growth kinetics.
\newblock {\em Phys. Rev. B}, 50:15752--15756, 1994.

\bibitem{Chong2012}
E.~K. Chong and S.~H. Zak.
\newblock {\em An Introduction to Optimization}.
\newblock John Wiley and Sons, Inc., New Jersey, 4 edition, 2012.

\bibitem{libo2018}
S.~B. Dai, B.~Li, and J.~F. Lu.
\newblock Convergence of phase-field free energy and boundary force for
  molecular solvation.
\newblock {\em Arch. Rational Mech. Anal.}, 227:105--147, 2018.

\bibitem{Selim2009}
M.~Elsey, S.~Esedoglu, and P.~Smereka.
\newblock Diffusion generated motion for grain growth in two and three
  dimensions.
\newblock {\em J. Comput. Phys.}, 228:8015--8033, 2009.

\bibitem{Selim2013}
M.~Elsey, S.~Esedoglu, and P.~Smereka.
\newblock Simulations of anisotropic grain growth: Efficient algorithms and
  misorientation distributions.
\newblock {\em Acta Mater.}, 61:2033--2043, 2013.

\bibitem{Liu2019gb2}
Y.~Epshteyn, C.~Liu, and M.~Mizuno.
\newblock Large time asymptotic behavior of grain boundaries motion with
  dynamic lattice misorientations and with triple junctions drag.
\newblock {\em arXiv:1910.08022}, 2019.

\bibitem{Liu2019gb}
Y.~Epshteyn, C.~Liu, and M.~Mizuno.
\newblock Motion of grain boundaries with dynamic lattice misorientations and
  with triple junctions drag.
\newblock {\em arXiv:1903.11512}, 2019.

\bibitem{Selim2016}
S.~Esedoglu.
\newblock Grain size distribution under simultaneous grain boundary migration
  and grain rotation in two dimensions.
\newblock {\em Comput. Mater. Sci.}, 121:209--216, 2016.

\bibitem{feng2003}
X.~B. Feng and A.~Prohl.
\newblock Numerical analysis of the {Allen-Cahn} equation and approximation for
  mean curvature flows.
\newblock {\em Numer. Math.}, 94:33--65, 2003.

\bibitem{Frank}
F.~C. Frank.
\newblock The resultant content of dislocations in an arbitrary
  intercrystalline boundary.
\newblock In {\em Symposium on the plastic deformation of crystalline solids},
  pages 150--154. Office of Naval Research, Pittsburgh, 1950.

\bibitem{Harris19982623}
K.~E. Harris, V.~V. Singh, and A.~H. King.
\newblock Grain rotation in thin films of gold.
\newblock {\em Acta Mater.}, 46:2623 -- 2633, 1998.

\bibitem{Herring1951}
C.~Herring.
\newblock Surface tension as a motivation for sintering.
\newblock In W.~E. Kingston, editor, {\em The Physics of Powder Metallurgy},
  pages 143--179. McGraw-Hill, New York, 1951.

\bibitem{HL}
J.~P. Hirth and J.~Lothe.
\newblock {\em Theory of Dislocations}.
\newblock Wiley, New York, second edition, 1982.

\bibitem{Kazaryan2000}
A.~Kazaryan, Y.~Wang, S.~A. Dregia, and B.~R. Patton.
\newblock Generalized phase-field model for computer simulation of grain growth
  in anisotropic systems.
\newblock {\em Phys. Rev. B}, 61:14275--14278, 2000.

\bibitem{liuchun2001}
D.~Kinderlehrer and C.~Liu.
\newblock Evolution of grain boundaries.
\newblock {\em Math. Models Methods Appl. Sci.}, 4:713--729, 2001.

\bibitem{kinderlehrer2006}
D.~Kinderlehrer, I.~Livshits, and S.~Tasan.
\newblock A variational approach to modeling and simulation of grain growth.
\newblock {\em SIAM J. Sci. Comput.}, 28:1694--1715, 2006.

\bibitem{Kobayashi2000}
R.~Kobayashi, J.~A. Warren, and W.~C. Carter.
\newblock A continuum model of grain boundaries.
\newblock {\em Phys. D}, 140:141--150, 2000.

\bibitem{Srolovitz2010}
E.~A. Lazar, R.~D. MacPherson, and D.~J. Srolovitz.
\newblock A more accurate two-dimensional grain growth algorithm.
\newblock {\em Acta Mater.}, 58:364--372, 2010.

\bibitem{Li1953223}
C.~H. Li, E.~H. Edwards, J.~Washburn, and E.~R. Parker.
\newblock Stress-induced movement of crystal boundaries.
\newblock {\em Acta Metall.}, 1:223--229, 1953.

\bibitem{Li1962}
J.~C.~M. Li.
\newblock Possibility of subgrain rotation during recrystallization.
\newblock {\em J. Appl. Phys.}, 33:2958--2965, 1962.

\bibitem{Voorhees2016264}
K.~McReynolds, K.~A Wu, and P.~Voorhees.
\newblock Grain growth and grain translation in crystals.
\newblock {\em Acta Mater.}, 120:264--272, 2016.

\bibitem{Molodov2007}
D.~A. Molodov, V.~A. Ivanov, and G.~Gottstein.
\newblock Low angle tilt boundary migration coupled to shear deformation.
\newblock {\em Acta Mater.}, 55:1843--1848, 2007.

\bibitem{Mullins1956}
W.~Mullins.
\newblock Two-dimensional motion of idealized grain boundaries.
\newblock {\em J. Appl. Phys.}, 27:900--904, 1956.

\bibitem{Rath2007}
B.~B. Rath, M.~Winning, and J.~C.~M. Li.
\newblock Coupling between grain growth and grain rotation.
\newblock {\em Appl. Phys. Lett.}, 90:161915, 2007.

\bibitem{ReadShockley1950}
W.~T. Read and W.~Shockley.
\newblock Dislocation models of crystal grain boundaries.
\newblock {\em Phys. Rev.}, 75:275--289, 1950.

\bibitem{Voigt2018}
M.~Salvalaglio, R.~Backofen, K.~R. Elder, and Axel Voigt.
\newblock Defects at grain boundaries: A coarse-grained, three-dimensional
  description by the amplitude expansion of the phase-field crystal model.
\newblock {\em Phys. Rev. Mater.}, 2:053804, 2018.

\bibitem{Shewmon1966}
P.~G. Shewmon.
\newblock Energy and structure of grain boundaries.
\newblock In H.~Margolin, editor, {\em Recrystallization, grain growth and
  textures}, pages 165--199. American Society of Metals, Metals Park, 1966.

\bibitem{Cahn20021}
S.~G. Srinivasan and J.~W. Cahn.
\newblock Challenging some free-energy reduction criteria for grain growth.
\newblock In S.~Ankem, C.~S. Pande, I.~Ovid'ko, and S.~Ranganathan, editors,
  {\em Science and Technology of Interfaces}, pages 3--14. TMS, Seattle, 2002.

\bibitem{Sutton1995}
A.~P. Sutton and R.~W. Balluffi.
\newblock {\em Interfaces in Crystalline Materials}.
\newblock Clarendon Press, Oxford, 1995.

\bibitem{Taylor2007493}
J.~E. Taylor and J.~W. Cahn.
\newblock Shape accommodation of a rotating embedded crystal via a new
  variational formulation.
\newblock {\em Interfaces and Free Boundaries}, 9:493--512, 2007.

\bibitem{Thomas2017}
S.~L. Thomas, K.~T. Chen, J.~Han, P.~K. Purohit, and D.~J. Srolovitz.
\newblock Reconciling grain growth and shear-coupled grain boundary migration.
\newblock {\em preprint}, 2017.

\bibitem{Trautt20122407}
Z.~T. Trautt and Y.~Mishin.
\newblock Grain boundary migration and grain rotation studied by molecular
  dynamics.
\newblock {\em Acta Mater.}, 60:2407--2424, 2012.

\bibitem{Upmanyu2002}
M.~Upmanyu, G.~N. Hassold, A.~Kazaryan, E.~A. Holm, Y.~Wang, B.~Patton, and
  D.~J. Srolovitz.
\newblock Boundary mobility and energy anisotropy effects on microstructural
  evolution during grain growth.
\newblock {\em Interface Sci.}, 10:201--216, 2002.

\bibitem{Upmanyu1998}
M.~Upmanyu, R.~W. Smith, and D.~J. Srolovitz.
\newblock Atomistic simulation of curvature driven grain boundary migration.
\newblock {\em Interface Sci.}, 6:41--58, 1998.

\bibitem{Upmanyu2006}
M.~Upmanyu, D.~J. Srolovitz, A.~E. Lobkovsky, J.~A. Warren, and W.~C. Carter.
\newblock Simultaneous grain boundary migration and grain rotation.
\newblock {\em Acta Mater.}, 54:1707--1719, 2006.

\bibitem{Wei2019133}
C.~Z. Wei, S.~L. Thomas, J.~Han, Y.~Xiang, and D.~J. Srolovitz.
\newblock A continuum multi-disconnection-mode model for grain boundary motion.
\newblock {\em J. Mech. Phys. Solids}, 133:103731, 2019.

\bibitem{Wu2012407}
K.~A. Wu and P.~W. Voorhees.
\newblock Phase field crystal simulations of nanocrystalline grain growth in
  two dimensions.
\newblock {\em Acta Mater.}, 60:407--419, 2012.

\bibitem{Voorhees2017}
A.~Yamanaka, K.~McReynold, and P.~W. Voorhees.
\newblock Phase field crystal simulation of grain boundary motion, grain
  rotation and dislocation reactions in a bcc bicrystal.
\newblock {\em Acta Mater.}, 133:160--171, 2017.

\bibitem{Zhang2005}
H.~Zhang, M.~Upmanyu, and D.J. Srolovitz.
\newblock Curvature driven grain boundary migration in aluminum: molecular
  dynamics simulations.
\newblock {\em Acta Mater.}, 53:79--86, 2005.

\bibitem{du2009}
J.~Zhang and Q.~Du.
\newblock Numerical studies of discrete approximations to the {Allen–Cahn}
  equation in the sharp interface limit.
\newblock {\em SIAM J. Sci. Comput.}, 31:3042--3063, 2009.

\bibitem{Zhang2017}
L.~C. Zhang, Y.~J. Gu, and Y.~Xiang.
\newblock Energy of low angle grain boundaries based on continuum dislocation
  structure.
\newblock {\em Acta Mater.}, 126:11--24, 2017.

\bibitem{Zhang2017119}
L.~C. Zhang, J.~Han, Y.~Xiang, and D.~J. Srolovitz.
\newblock Equation of motion for a grain boundary.
\newblock {\em Phys. Rev. Lett.}, 119:246101, 2017.

\bibitem{ZhangXiang2018157}
L.~C. Zhang and Y.~Xiang.
\newblock Motion of grain boundaries incorporating dislocation structure.
\newblock {\em J. Mech. Phys. Solids}, 117:157--178, 2018.

\bibitem{Zhu2014175}
X.~H. Zhu and Y.~Xiang.
\newblock Continuum framework for dislocation structure, energy and dynamics of
  dislocation arrays and low angle grain boundaries.
\newblock {\em J. Mech. Phys. Solids}, 69:175 -- 194, 2014.

\end{thebibliography}
\end{document}